\documentclass[apj]{emulateapj}
\usepackage{natbib}
\usepackage{graphicx}
\usepackage[toc,page]{appendix}
\usepackage{natbib}
\usepackage{color}
\usepackage{threeparttable}
\usepackage{hyperref}
\usepackage{breakurl}
\usepackage{amsmath}
\begin{document}

\slugcomment{Accepted to be published in the Astrophysical Journal}
\title{Predicting Quiescence: The Dependence of Specific Star Formation Rate \\ on Galaxy Size and Central Density at 0.5$<$z$<$2.5}
\email{kwhitaker@astro.umass.edu}
\author{Katherine E. Whitaker\altaffilmark{1,2,10}, Rachel Bezanson\altaffilmark{3,10}, Pieter G. van Dokkum\altaffilmark{4},
	Marijn Franx\altaffilmark{5}, Arjen van der Wel\altaffilmark{6}, Gabriel Brammer\altaffilmark{7},
	Natascha M. F\"{o}rster-Schreiber\altaffilmark{8},
	Mauro Giavalisco\altaffilmark{1}, Ivo Labb\'{e}\altaffilmark{5}, Ivelina G. Momcheva\altaffilmark{7}, 
		Erica J. Nelson\altaffilmark{4,8}, Rosalind Skelton\altaffilmark{9}}	
\altaffiltext{1}{Department of Astronomy, University of Massachusetts, Amherst, MA 01003, USA}
\altaffiltext{2}{Department of Physics, University of Connecticut, Storrs, CT 06269, USA}
\altaffiltext{3}{Steward Observatory, Department of Astronomy, University of Arizona, AZ 85721, USA}
\altaffiltext{4}{Leiden Observatory, Leiden University, P.O. Box 9513, 2300 RA Leiden, The Netherlands}
\altaffiltext{5}{Department of Astronomy, Yale University, New Haven, CT 06520, USA}
\altaffiltext{6}{Max-Planck Institut für Astronomie, Königstuhl 17, D-69117, Heidelberg, Germany 0000-0002-5027-0135}
\altaffiltext{7}{Space Telescope Science Institute, Baltimore, MD 21218, USA}
\altaffiltext{8}{Max-Planck-Institut für extraterrestrische Physik, Giessenbachstrasse, D-85748 Garching, Germany}
\altaffiltext{9}{South African Astronomical Observatory, P.O. Box 9, Observatory, Cape Town, 7935, South Africa}
\altaffiltext{10}{Hubble Fellow}
\shortauthors{Whitaker et al.}
\shorttitle{Predicting Quiescence: The Dependence of sSFR on Galaxy Size and Central Density}

\begin{abstract}
In this paper, we investigate the relationship between star formation and structure,
using a mass-complete sample of 27,893 galaxies at 0.5$<$$z$$<$2.5 selected from 3D-HST.  
We confirm that star-forming galaxies are larger than quiescent galaxies at fixed stellar mass (M$_{\star}$). 
However, in contrast with some simulations, there is only a weak relation between 
star formation rate (SFR) and size within the star-forming population: when dividing
into quartiles based on residual offsets in SFR, 
we find that the sizes of star-forming galaxies in the lowest quartile are
0.27$\pm$0.06 dex smaller than the highest quartile.  
We show that 50\% of star formation in galaxies at 
fixed M$_{\star}$ takes place within a narrow range of sizes (0.26 dex).  Taken together,
these results suggest that there is an abrupt cessation of star formation after galaxies
attain particular structural properties.  Confirming earlier results,
we find that central stellar density within a 1 kpc fixed physical radius is the 
key parameter connecting 
galaxy morphology and star formation histories: galaxies with high central densities  
are red and have increasingly lower SFR/M$_{\star}$, whereas galaxies with low 
central densities are blue and 
have a roughly constant (higher) SFR/M$_{\star}$ at a given redshift.  
We find remarkably little scatter in the average trends and a strong evolution of $>$0.5 dex 
in the central density threshold correlated with quiescence from $z$$\sim$0.7-2.0.
Neither a compact size nor high-$n$ are sufficient to assess the likelihood of 
quiescence for the average galaxy; rather, the combination of these two parameters
together with M$_{\star}$ results in a unique quenching threshold in central density/velocity.
\end{abstract}


\section{Introduction}
\label{sec:intro}

Despite decades of deep and wide extragalactic surveys, we still 
do not understand the astrophysics behind the empirical relationship linking the star formation
histories of galaxies and their morphologies. Observations show
that galaxies with evolved stellar populations, so-called ``quiescent''
galaxies, have significantly smaller sizes and more concentrated light profiles
than actively star-forming galaxies with a similar stellar mass and redshift
\citep[e.g.,][]{Shen03, Trujillo07, Cimatti08, Kriek09b, Williams10, Wuyts11b, vanderWel14}.
Although we know that galaxies must shut down their star formation
and migrate from the star-forming to quiescent population, there is much to
be learned about the physical process(es) that are primarily responsible for
this structural evolution and the quenching of star formation. 

One way to study the connection between this bimodal population of galaxies
is through correlations between specific star formation rate (sSFR$\equiv$SFR/M$_{\star}$)
and parameters describing various physical properties of galaxies, such as stellar mass 
\citep[e.g.,][]{Whitaker14b,Schreiber15}, surface density \citep[e.g.,][]{Franx08,Barro13},
bulge mass \citep[e.g.,][]{Lang14,Bluck14,Schreiber16}, or environment \citep[e.g.][]{Elbaz07}.
The inverse of the sSFR defines a timescale for the formation of the stellar population of a galaxy,
where lower sSFRs correspond to older stellar populations for a constant or single burst star formation history.  
In this sense, sSFR is a relatively straight forward diagnostic of quiescence that can be 
directly linked to other physical properties of galaxies.

With a sample of galaxies selected from the Sloan Digital Sky Survey (SDSS), 
\citet{Brinchmann04} were the first to show that there is a turnover in 
the sSFR of galaxies at higher stellar surface mass densities \citep[also studied in the context
of a turnover in D$_{\mathrm{n}}$(4000) by][]{Kauffmann03c}.  The redshift
evolution of this correlation was later presented in \citet{Franx08} \citep[see also][]{Maier09}.
Both works identified a threshold surface 
density at each redshift interval: below this threshold the sSFRs are
high with little variation, and above the threshold density galaxies
have low sSFRs. \citet{Franx08} reported that the density threshold increases with redshift, at least
out to $z$=3.
As stellar density and velocity dispersion are closely related \citep[e.g.,][]{Wake12, Fang13},
observations therefore indicate that galaxies are statistically more likely to be quiescent 
once they have surpassed a threshold in either density or velocity dispersion.
Studies of early-type galaxies at $z$$\sim$0 further show that at fixed stellar mass 
velocity dispersion is strongly correlated with other physical properties: galaxies with increased velocity dispersion 
and thereby more compact sizes are on average older, 
more metal-rich, with lower molecular gas fractions, and more alpha-enhanced than their larger, 
lower velocity dispersion counterparts \citep{Thomas05, Cappellari13, McDermid15}.
 
\citet{Bezanson09} showed that distant compact galaxies have similar densities to the central 
regions of these local early-type galaxies by comparing their average stellar density profiles within a 
constant physical radius of 1 kpc \citep[see also][]{vanDokkum10,Saracco12,vanDokkum14,Tacchella15b}.
This study was the first to present a plausible link between
 these high redshift galaxies and where they end up in the local universe, but it is yet unclear what
 causes the quenching in the first place.  This work does however suggest that it may be more robust to 
 define a quenching threshold in surface density within the \emph{central} 1 kpc, 
as opposed to the half-light radius.
\citet{Fang13} find that this central density threshold increases with stellar mass through a
study of the correlation between galaxy structure and the quenching of 
star formation using a sample of SDSS central galaxies.  Furthermore, studies that push
the analysis of central density out to $z$=3 corroborate \citep[e.g.,][]{Cheung12,Saracco12,Barro13,Barro15}, 
supporting the idea that the innermost structure of galaxies is most physically linked with quenching.  
Where the earlier work of 
\citet{Franx08} found an evolving effective surface density threshold with redshift, \citet{Barro15} 
do not find a strong redshift evolution in the central surface density threshold. 
However, as there still exist star-forming galaxies above this quenching threshold, results in the literature conclude
that a dense bulge is a necessary but insufficient condition to fully quench galaxies \citep[see also][]{Bell12}. 

While most studies have focused on the the stellar mass dependence of the central density alone, 
it is perhaps unsurprising that there is a tight correlation: the central density is a biproduct of the
combined stellar mass and light profile.  The key comparison instead should be with sSFR, normalizing out the stellar
mass dependence of SFR \citep[as also studied in][]{Barro15}, where total sSFRs can be measured largely 
independent of the central density.
While the dynamic range in stellar mass enabled by the deep high resolution near-infrared (NIR)
imaging from the Cosmic Assembly Near-IR Deep Extragalactic Legacy 
Survey \citep[CANDELS;][]{Grogin11, Koekemoer11} improves dramatically from earlier multi-wavelength
extragalactic surveys \citep[e.g.,][]{Wuyts08, Whitaker12b}, the depth of the Spitzer/MIPS 24$\mu$m imaging used
to derive the IR SFR indicator has remained unchanged.  To therefore leverage the full range in stellar mass
and galaxy structure probed by these \emph{Hubble Space Telescope} legacy programs,
we must perform detailed stacking analyses of the 24$\mu$m imaging to probe the SFR
properties of the complete unbiased sample of galaxies using a single robust SFR indicator \citep[e.g.,][]{Whitaker15}.

By combining the high resolution photometry from CANDELS with the accurate spectroscopic information provided 
by the 3D-HST treasury program \citep{Brammer12,Momcheva15} and a stacking analysis of the unobscured (UV) and obscured (IR) SFRs,
we are in a unique position to perform a census across most of cosmic time of the simultaneous 
evolution of galaxy structure and star formation.
While earlier results from this treasury dataset have shown that all quiescent galaxies have a dense 
stellar core, and the formation of such cores is a requirement for quenching \citep{vanDokkum14, Whitaker15, Barro15}, 
there are several open questions that we aim to answer in this paper.  
Specifically, (1) how does star formation rate depend on galaxy size?, 
(2) is there a preferential galaxy size scale where star formation occurs?, 
(3) is there a physical parameter that will uniquely predict quiescence?, and 
(4) does the quenching threshold in surface density and velocity evolve with redshift?    
There are a few differences that together separate the present analysis from earlier studies: 
the inclusion of accurate grism redshifts from 3D-HST improve the stellar population parameters, 
we derive the three dimensional deprojected central density and circular velocity instead
of the surface density, and we stack the 24$\mu$m imaging to robustly measure total SFRs for
more extended, lower stellar mass, or low SFR galaxies.  

The paper is outlined as follows. In Section \ref{sec:data}, we introduce the data and sample
selection, describing the details of the stellar masses, redshifts, rest-frame colors, structural
parameters, total star formation rates, central densities and circular velocities used herein.
We present the correlations between galaxy size, stellar mass and sSFR for the overall population 
in Section~\ref{sec:size}.  In Section~\ref{sec:sizescale},
we determine at which galaxy size scale the most star formation occurs from $z$=0.5 to $z$=2.5.
And in Section~\ref{sec:sfalone}, we proceed to analyze the residual offsets in SFR and size 
for star-forming galaxies alone once removing the well known correlations between
log(SFR)--log(M$_{\star}$) and log($r_{e}$)--log(M$_{\star}$).  
In the second half of the paper, we explore the physical parameters that best predict quiescence.
First, we consider the role of galaxy size and S\'{e}rsic index in predicting quiescence in Section~\ref{sec:sersic}.
Next, in Section~\ref{sec:density}, we study the dependence of sSFR on stellar mass density,
parameterizing the redshift evolution in Section~\ref{sec:evolution}, and the density and velocity quenching thresholds
in Section~\ref{sec:evolve_quiescence}.
As this paper touches on a relatively wide range of topics, we integrate the discussion and implications 
of the results throughout the relevant sections, as well as further discussion in Section~\ref{sec:discussion}.  
While we choose to place these empirical results in the context of current theoretical models, 
we note that many of the correlations that we discuss can be interpreted in a different way \citep[e.g.,][]{Lilly16, Abramson16}.  
We caution that it is yet unclear if there is truly an evolutionary sequence causally linking galaxy 
structure with star formation.
We conclude the paper with a summary in Section~\ref{sec:summary} of the results presented herein,
in the context of current and future studies of galaxy formation and evolution.

In this paper, we use a \citet{Chabrier} initial mass function (IMF) and assume a $\Lambda$CDM 
cosmology with $\Omega_{\mathrm{M}}=0.3$, 
$\Omega_{\Lambda}=0.7$, and $\mathrm{H_{0}}=70$ km s$^{-1}$ Mpc$^{-1}$. All magnitudes are given in the AB system.

\section{Data and Sample Selection}
\label{sec:data}

\subsection{Stellar Masses, Redshifts and Rest-frame Colors}
We use the exquisite \emph{HST} Wide Field Camera 3 (WFC3) multi-wavelength photometric and spectroscopic datasets of five 
well-studied extragalactic 
fields through the CANDELS and 3D-HST surveys.  
Using stellar masses, redshifts, and rest-frame colors from the 3D-HST 0.3--8$\mu$m photometric 
catalogs \citep[see][for full details]{Skelton14}, we select samples in three redshift 
intervals of 0.5$<$$z$$<$1.0, 1.0$<$$z$$<$1.5, and 1.5$<$$z$$<$2.5: 11266, 9553, and 7791 galaxies above
the stellar mass limits for star-forming galaxies.  When splitting the sample into sub-populations and 
accounting for stellar mass limits, our sample is reduced
further to 9694 (1192), 8643 (705), and 
6893 (766) star-forming (quiescent) galaxies greater than stellar mass limits 
of $\log\mathrm{M_{\star}/M_{\odot}}=8.6 (9.0)$, 8.8 (9.4), and 9.4 (10.0), respectively.  
The galaxies are 
defined to be either star-forming or quiescent based on their rest-frame $U$--$V$ and $V$--$J$ colors, following
the definition of \citet{Whitaker12a}.  
We have identified and removed luminous active galactic nuclei (AGNs) using 
the \emph{Spitzer}/IRAC color selections presented in \citet{Donley12}, as they may have significant
contamination of their IR SFR; only 2\% of the
sample were removed as AGN candidates. 
The total sample comprises 27,893 galaxies at 0.5$<$$z$$<$2.5.

We only analyze data where we are mass-complete for star-forming galaxies.  
The lower bounds of the stellar mass limits correspond to the mass-completeness limits down to which \citet{vanderWel14}
can determine structural parameters for star-forming and quiescent galaxies with good fidelity.  
The values for star-forming galaxies are in agreement with the mass-completeness limits presented 
in \citet{Tal14}, which are determined by 
comparing object detection in CANDELS/deep with a re-combined subset of the exposures which 
reach the depth of CANDELS/wide.  
We indicate the stellar mass limits from \citet{vanderWel14} to which we can trust the 
star-forming and quiescent structural measurements within Figures~\ref{fig:size} and \ref{fig:sfuniverse}.
We additionally correct the stellar masses of star-forming galaxies 
for contamination of the broadband fluxes from emission-lines using the values presented in 
Appendix A of \citet{Whitaker14b}.  
These corrections only begin to become significant at $\log\mathrm{M_{\star}/M_{\odot}}<9.5$ and $z>1.5$.

Where available, we combine the spectral energy distributions (SEDs) with low-resolution 
HST/WFC3 G141 grism spectroscopy to derive
grism redshifts with 0.3\% accuracy \citep{Brammer12}.                      
\citet{Momcheva15} present the full details of the 3D-HST grism data reduction and redshift analysis.
We select the ``best'' redshift to be the spectroscopic redshift, grism redshift or the photometric 
redshift, in this ranked order depending on the availability.
Photometric redshifts comprise 52\% (57\%) of the $0.5<z<1.5$ ($1.5<z<2.5$) sample, 
while 39\% (40\%) have grism redshifts and 9\% (2\%) spectroscopic redshifts.

\subsection{Structural Parameters}
\label{sec:structure}
Size and S\'{e}rsic indices used herein are measured from deep HST/WFC3 $J_{\mathrm{F125W}}$ 
and $H_{\mathrm{F160W}}$ photometry, as presented in \citet{vanderWel12, vanderWel14}.
The structural measurements are parameterized profile fits which implicitly 
take into account the HST/WFC3 point spread function at the time of the measurement.  
These measurements therefore do not require systematic corrections of more than a few percent 
\citep[see Section 2.5 in][]{vanderWel12}, and represent the size (and S\'{e}rsic index) 
distribution with good fidelity across the examined redshift range.
The effective radius in each filter is defined to be 
the semi-major axis of the ellipse that contains half of the total flux of the best-fitting S\'{e}rsic model. 
\citet{vanderWel14} parameterize the effective radius as a simple function of redshift and stellar mass
(their Equations 1 and 2).  Following Equation 2 in \citet{vanderWel14}, the rest-frame 5000\AA\ effective radius 
for galaxies with $z$$<$1.5 is measured from the $J_{\mathrm{F125W}}$ effective radius, whereas
$H_{\mathrm{F160W}}$ is used at $z$$>$1.5.  Similarly, we adopt
the S\'{e}rsic indices measured from the $J_{\mathrm{F125W}}$ photometry at $0.5<z<1.5$ and 
$H_{\mathrm{F160W}}$ at $1.5<z<2.5$ \citep[see][for details]{Whitaker15} for the central density
measurement in Section~\ref{sec:centraldensity}.  The details of the error analysis on $r_{e}$ and $n$ 
are presented in \citet{vanderWel12}.  

\subsection{Total Star Formation Rates}
Total star formation rates are derived from median stacks of Spitzer/MIPS 24$\mu$m photometry, 
following the procedure detailed in \citet{Whitaker14b}. The Spitzer/MIPS 24$\mu$m 
images in the AEGIS field 
are provided by the Far-Infrared Deep Extragalactic Legacy (FIDEL) survey \citep{Dickinson07}, COSMOS from
the S-COSMOS survey \citep{Sanders07}, GOODS-N and GOODS-S from \citet{Dickinson03}, and 
UDS from the Spitzer UKIDSS Ultra Deep 
Survey\footnote{\url{http://irsa.ipac.caltech.edu/data/SPITZER/SpUDS/}}(SpUDS; PI: J. Dunlop).  

Briefly, the analysis code uses a high-resolution $J_{\mathrm{F125W}}$+$H_{\mathrm{F140W}}$+$H_{\mathrm{F160W}}$ 
detection image as a prior to model the contributions from neighboring blended sources in the lower resolution 
MIPS 24$\mu$m image.  All galaxies are ``cleaned'' of the contaminating flux of the neighboring sources before
stacking. We refer the reader to Section 3 of \citet{Whitaker14b} for the full details of the MIPS 24$\mu$m stacking
analyses. The SFRs derived for quiescent galaxies herein are likely upper limits, as the 24$\mu$m technique
tends to overestimate the SFRs for galaxies with $\log\mathrm{sSFR}$$<$-10 yr$^{-1}$ \citep{Hayward14, Utomo14, Fumagalli14}.  We note 
that the UV+IR SFR technique is generally robust for star-forming galaxies.
We derive uncertainties in the average star formation rates from 50 Monte Carlo bootstrap
simulations of the stacking analyses.  The error in the mean is therefore the width of the resulting distribution
of star formation rates divided by the square root of the number of galaxies in each bin. 

We choose to use the Spitzer/MIPS 24$\mu$m IR SFR because of the resolution and depth of the observations, 
and to mitigate 
systematic uncertainties when combining different SFR indicators.   However, we note that the observed 
24$\mu$m samples major spectral features arising from polycyclic aromatic hydrocarbons (PAHs).  
Despite complications from these PAH features, \citet{Wuyts11a} demonstrate that the luminosity-independent 
conversion from 24$\mu$m to the bolometric IR luminosity used here yields estimates that are
in good median agreement with measurements from Herschel/PACS 
photometry \citep[see also][and B. Lee et al. in prep]{Tomczak16}.  
When combining the 24$\mu$m IR SFR indicator
with the rest-frame UV, we can therefore successfully recover the average total amount of star formation in galaxies.
While considering the average correlations and including a bootstrap error analysis will reduce the potential 
noise in these measurements, 
we cannot rule out that there exist biases due to the physical conditions of the dust and the 
star formation itself that are incredibly difficult to quantify.

\subsection{Central Density}
\label{sec:centraldensity}

Given the observed projected two dimensional surface density, the surface brightness distribution
can be deprojected to obtain a three dimensional light distribution if we assume spherical symmetry.
We derive this three-dimensional density profile from the best-fit structural parameters to 
the intensity profiles of the individual galaxies described
in Section~\ref{sec:structure}.  The projected luminosity within a projected radius is 
described in \citet{Ciotti91}, assuming isotropic spherical galaxies with surface luminosity
profiles following the S\'{e}rsic profile.  We adopt the asymptotic approximation for the term $b_{n}$ from
\citet{Ciotti99}.  This asymptotic expansion is truncated to the first four terms and is accurate to
$6\times10^{-7}$ for exponential disks ($n$=1) and $10^{-7}$ for a de Vaucouleur profile ($n$=4).
The approximation for $b_{n}$ presented in \citet{Ciotti99} performs much better than previous 
formulae \citep[e.g.,][and others]{Ciotti91}.
However, despite the accuracy of the asymptotic expansion, we note that this methodology may 
lead to errors for galaxies that are far from the assumed spherical symmetry, in particular for flat disks.
We return to this issue in Section~\ref{sec:density}.
 
Following the equations summarized in Section 2.2 of \citet{Bezanson09}, we perform an Abel Transform to
deproject the circularized, three-dimensional light profile.  Assuming mass follows the light
and there are no strong color gradients, the total luminosity is converted
to a total stellar mass using the stellar masses presented in \citet{Skelton14}. These stellar 
masses (labeled $M_{\mathrm{phot}}$ below) are derived by modeling the spectral energy distributions and 
correspond to the final 3D-HST data release\footnote{\url{http://3dhst.research.yale.edu/Data.php}}.
Following \citet{vanDokkum14}, we apply a small correction to these stellar masses to take into account 
the difference between the total magnitude in the photometric catalog and the total magnitude
implied by the S\'{e}rsic fit \citep[see][]{Taylor10}.  On average, this correction is 1.03$\pm$0.11.
The central density is therefore calculated by numerically integrating the following equation:

\begin{equation}
\rho_{1} (r<1~\mathrm{kpc}) = \frac{\int_{0}^{1~\mathrm{kpc}}\rho(r)r^{2}dr}{\int_{0}^{\infty}\rho(r)r^{2}dr}\frac{L_{\mathrm{model}}}{L_{\mathrm{phot}}}\frac{M_{\mathrm{phot}}}{\frac{4}{3}\pi (1~\mathrm{kpc})^{3}}
\label{eq:rho1}
\end{equation}

\noindent where $L_{\mathrm{phot}}$ is the total, aperture-corrected luminosity of the galaxy from the 3D-HST
catalogs in the filter corresponding to the S\'{e}rsic profile measurement 
(e.g., $J_{\mathrm{F125W}}$ or $H_{\mathrm{F160W}}$).
$L_{\mathrm{model}}$ is the total luminosity as measured from integrating the best-fit S\'{e}rsic profile.
In 10\% of the cases, the numerical integration does not converge and these unreliable
measurements are removed from the subsequent analysis.  We find that these galaxies generally have 
Gaussian profiles with $n$=0.45$\pm$0.14, larger than average sizes with $r_{e}$=3.4$\pm$1.5 kpc, and low stellar masses
within $\lesssim$0.5 dex of the stellar mass limits; these galaxies represent precisely the population one might expect to fail.  
While we are effectively using the same formalism as \citet{vanDokkum14}, who
derive the ``core'' mass within the central 1 kpc, we instead parameterize the central
density and circular velocity within 1 kpc to facilitate comparisons to earlier results by \citet{Franx08}.
We note that these derived parameters are essentially equivalent, modulo constant factors, where
M$_{1}$$\propto$$\rho_{1}$$\propto$$v_{\mathrm{circ},1}^{2}$. 

\begin{figure*}[t]
\leavevmode
\centering
\includegraphics[width=0.83\linewidth]{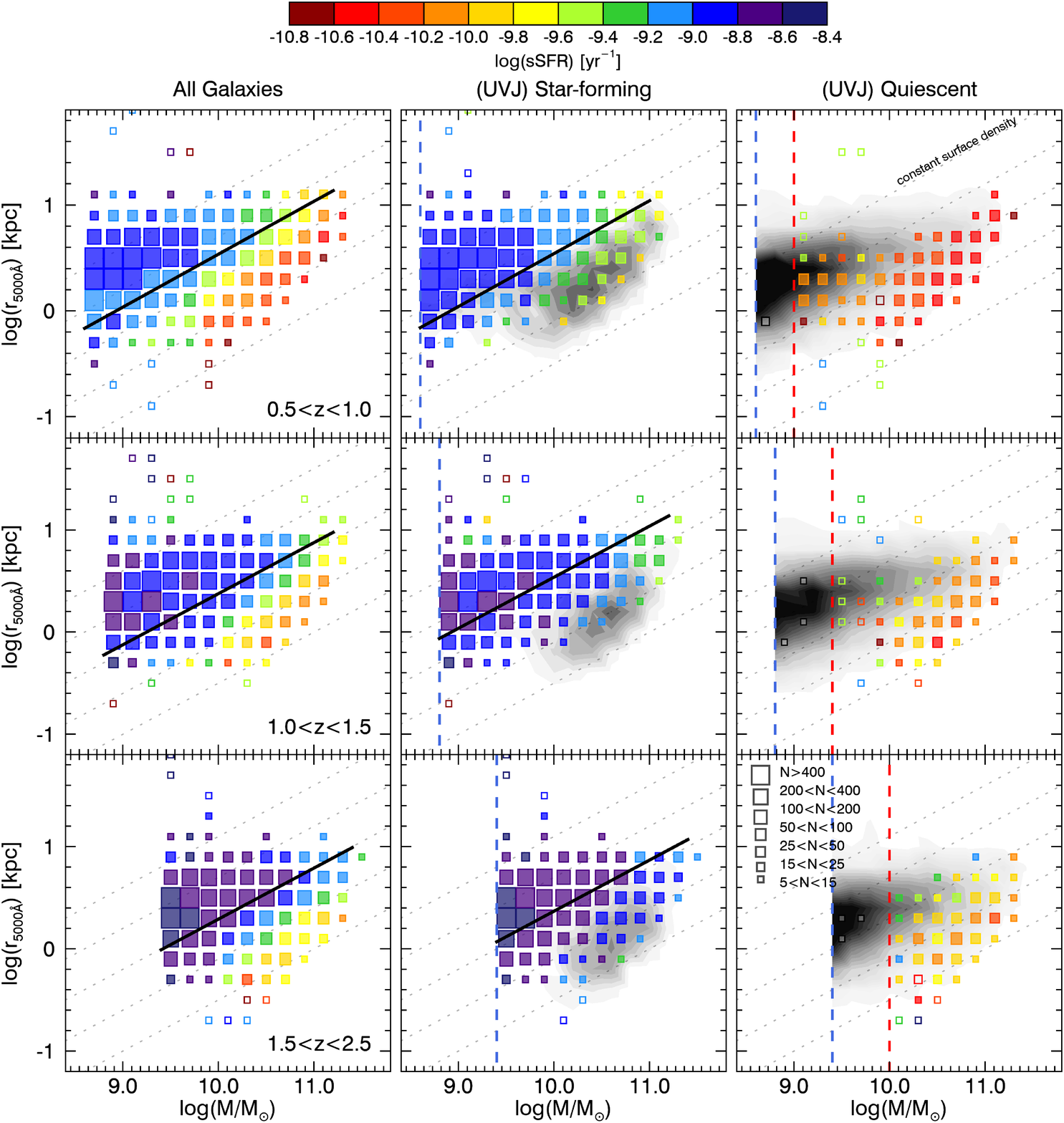}
\caption{Rest-frame 5000\AA\ size of galaxies as a function of stellar mass, color-coded by the sSFRs derived
from UV+IR median stacking analyses in 0.2 dex bins of $\log\mathrm{M_{\star}/M_{\odot}}$ and $\log\mathrm{r_{5000\AA}}$.
The size of the symbol depends on the number of galaxies that enter each bin.                         
The vertical dashed lines correspond to the stellar mass limits down to which the structural
parameters can be trusted for star-forming (blue) and
quiescent (red) populations.  The black dotted lines correspond to lines of constant 
surface density, stellar mass per unit area, with the solid line corresponding to the characteristic
central density measured in Figure~\ref{fig:sigma1_ssfr}. } 
\label{fig:size}
\end{figure*}

Uncertainties in $\rho_{1}$ will originate from how well we can measure $r_{e}$  and $n$ 
for each individual galaxy, and to first order this uncertainty depends on S/N.  Using the CANDELS 
wide $H_{\mathrm{F160W}}$ imaging, \citet{vanderWel12} show that the parameter $r_{e}$ can be inferred 
with a systematic uncertainty of 10\% or better for galaxies brighter than $H_{\mathrm{F160W}}\sim24$, 
whereas $n$ can be measured at the same level of accuracy for galaxies brighter than $H_{\mathrm{F160W}}\sim23$.  
We adopt the corresponding limits in stellar mass for reliable $r_{e}$ for the subsequent analysis.  
We note however that these stellar mass limits are roughly 0.5 dex lower than those for reliable $n$, 
as $n$ is more challenging to constrain.  We return to this issue in Section~\ref{sec:evolution}.

To quantify the errors on the central density, we perform 50 bootstrap simulations of this numerical integration, 
perturbing $r_{e}$ by pseudo random offsets drawn from a Gaussian distribution with a standard deviation equal 
to the respective 1$\sigma$ errors, as calculated by \citet{vanderWel12}. As the errors on $r_{e}$ are strongly 
correlated with both $n$ and magnitude/stellar mass \citep{Haussler07, Guo09, Bruce12, vanderWel12}, we use 
the following equations, $\Delta(\log r_{e}) = -0.25\Delta(\log M_{\star}))$ and 
$\Delta(\log n) = -0.27\Delta(\log M_{\star}))$ \citep[estimated from Figure 7 in][]{vanderWel12}, to 
derive the offsets in $n$ and M$_{\star}$ given $\Delta(\log r_{e})$ for each bootstrap iteration of 
the numerical integration. The width of the resulting distribution of central densities is taken as the 
error on each individual measurement of $\rho_{1}$. The error on the average central density for each bin is 
then the square root of the sum of the errors in quadrature divided by the number of galaxies in the bin.

\subsection{Central Circular Velocity}
\label{sec:vcirc}

If we balance the gravitational force acting on the mass enclosed within the central 1 kpc with the
centrifugal force, assuming spherical symmetry,
the circular velocity of a test particle at radius $r$=1 kpc will be,

\begin{equation}
v_{\mathrm{circ},1} (r<1~\mathrm{kpc}) = \sqrt{\frac{GM(r<1~\mathrm{kpc})}{1~\mathrm{kpc}}} = \sqrt{\frac{4}{3}  \pi G \rho_{1}}
\end{equation}

\noindent where $G$ is the gravitational constant and equal to
4.302$\times$10$^{-6}$ kpc M$_{\odot}^{-1}$ (km/s)$^{2}$,
and the stellar mass enclosed within 
the central 1 kpc sphere is determined from Equation~\ref{eq:rho1}.  
The central circular velocity is a factor of $\sqrt{2}$ greater than the velocity dispersion,
which both \citet{Franx08} and \citet{vanDokkum15} study.
We again caution that we must be careful to test that the results presented herein are not
driven by flat galaxies, which deviate from the assumption of spherical symmetry.
We return to this issue in Section~\ref{sec:density}.

\section{The Dependence of Star Formation Rate on Galaxy Size}

\subsection{How does star formation rate depend on galaxy size?}
\label{sec:size}

In Figure~\ref{fig:size}, we present the rest-frame 5000\AA\ size of galaxies as a function of their
stellar mass in three redshift intervals, 0.5$<$$z$$<$1.0, 1.0$<$$z$$<$1.5, and 1.5$<$$z$$<$2.5.  The data are
grouped into 0.2 dex bins of logarithmic stellar mass and size.  The size of the symbol represents
the number of galaxies that populate that area in parameter space.  The symbols are color-coded by the 
sSFR derived
from the UV+IR median stacking analysis, with open symbols signifying upper limits.  Whereas all galaxies are 
included in the left panels, the middle and right panels are separated into the star-forming and quiescent
populations as determined from rest-frame U-V and V-J colors.  The middle
and right panels of Figure~\ref{fig:size} additionally show the size-mass
contours in greyscale of the opposite population, quiescent and star-forming respectively.
We alternatively show a similar figure in the Appendix, which is instead 
color-coded by the deviation from the average
log(SFR)-log(M$_{\star}$) relation from \citet{Whitaker14b}.

For the overall galaxy population (Figure~\ref{fig:size}, left), 
galaxies with smaller sizes at fixed stellar mass are forming stars at lower rates.
We confirm the earlier results of \citet{vanderWel14} and numerous others, 
who show that the compact sizes of more massive quiescent galaxies are offset from the average size of 
star-forming galaxies by factors of approximately four at fixed stellar mass at least out to $z$$\sim$2.5.
Here we show that these compact galaxies indeed have low sSFRs \citep[see also][]{vanDokkum15}.  
As \citet{Fumagalli14} point out, the true SFRs of quiescent galaxies may be even lower as 
the mid-IR flux density can originate from processes unrelated to ongoing star formation, such as 
cirrus dust heated by old stellar populations and circumstellar dust. 
The UV+IR SFRs derived for quiescent galaxies herein are therefore likely upper limits (right panels in Figure~\ref{fig:size}), 
with the effect setting in for 24$\mu$m-derived SFRs with $\log\mathrm{sSFR}$$<$-10 yr$^{-1}$
\citep[e.g.,][]{Utomo14}.  Accounting for the overestimation of the quiescent SFRs 
by treating the SFRs measured for $\log\mathrm{sSFR}$$<$-10 yr$^{-1}$ as upper limits will 
only serve to accentuate the trends between galaxy size and SFR for the overall population.

In the right most panels of Figure~\ref{fig:size}, we see a flattening of the galaxy size-mass 
relation for quiescent galaxies at low stellar masses ($<$10$^{10}$ M$_{\odot}$), 
similar to \citet{Cappellari13} and \citet{Norris14}.  
This is most evident at 0.5$<$$z$$<$1.0,
where the stellar mass limits imposed by the structural measurements extend down to 10$^{9}$ M$_{\odot}$.
This flattening is not likely due to to an inability to measure small galaxy sizes due to the 
HST resolution limit
within the mass/magnitude limits we adopt, as \citet{vanderWel12} show that the sizes of small galaxies are 
not overestimated if they have sufficient S/N. \citet{vanderWel12} compared measurements from data with different depths 
(CANDELS deep vs. wide), as well as simulated S\'{e}rsic profiles.  The former analysis will quantify 
random errors and take into account that galaxies aren't necessarily well described by Sersic profiles;  
the latter analysis is useful for understanding systematic effects under the assumption that galaxies 
are well described by S\'{e}rsic profiles.  There is strong evidence suggesting that the S\'{e}rsic indices 
we measure from CANDELS data are robust, as we don’t appear to be missing light at large radii due to 
lack of depth \citep{vanderWel08,Szomoru10,Szomoru13}.
These low-mass quiescent galaxies therefore appear to have sizes similar to the bulk of the star-forming population
at that epoch, as well as slightly higher sSFRs than more massive quenched galaxies.
Simulations also show this flattening in the 
slope of the size-mass relation at stellar masses below 10$^{10}$ M$_{\odot}$ amongst quiescent galaxies 
\citep[e.g.,][]{Shankar14, Furlong15}.
These results hint at a more gradual quenching of star formation, perhaps due to a depleted
gas supply. These trends are likely not driven by environmental effects: results from \citet{Huertas-Company13} find no 
significant environmental dependence of the sizes of central and satellite quiescent galaxies at fixed 
stellar mass at $z$$\sim$0.

\begin{figure*}[t!]
\leavevmode
\centering
\includegraphics[width=0.9\linewidth]{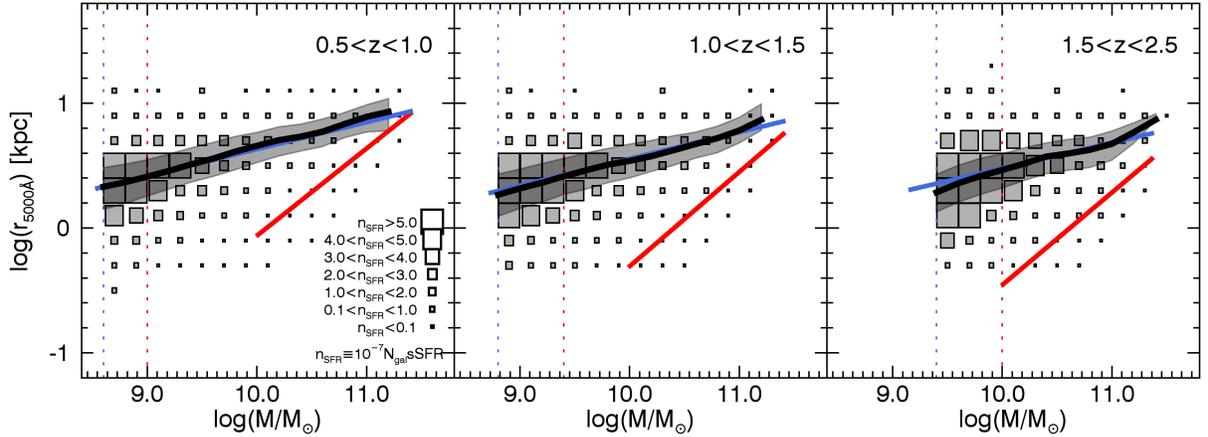}
\caption{The galaxy size-mass plane at 0.5$<$$z$$<$1.0, 1.0$<$$z$$<$1.5, and 1.5$<$$z$$<$2.5, with symbol sizes 
representing the total contribution to the star formation budget.  The size of each symbol is set by the 
number of galaxies within each 0.2 dex bin of $\log\mathrm{M_{\star}/M_{\odot}}$ and 
$\log\mathrm{r_{5000\AA}}$, multiplied by the median UV+IR specific SFR.  
The black line demarcates the 50th
percentile, signifying the size scale at which half of the star formation for a given stellar mass galaxy is occurring.  
The grey shaded region shows the 25th and 75th percentiles, showing that most of the star formation in the 
universe occurs within a relatively narrow range in galaxy sizes. The blue and red lines are the average size-mass
relations for star-forming and quiescent galaxies from \citet{vanderWel14}, respectively.}
\label{fig:sfuniverse}
\end{figure*}

\subsection{At what galaxy size scale does most star formation occur?}
\label{sec:sizescale}

With the present observations, it is interesting to consider on what galaxy size scale 
most of the stars in the Universe form.
In Figure~\ref{fig:sfuniverse}, the symbol size represents the product of the median sSFR for that
bin with the number of galaxies.  A symbol can be small either because there are few galaxies
that populate that parameter space and/or because those galaxies are not forming very many 
new stars on average.
As we take the median when deriving the UV+IR SFRs, star-forming galaxies far 
above the main ridgeline of the star formation sequence will not dominate the stacks.
The black line demarcates the 50th percentile, signifying the size scale at which half of the 
star formation for a given stellar mass occurs.  The grey shaded region then shows the 25th
and 75th percentile range, encompassing half of the star formation in the Universe.  
These percentiles were determined by rank ordering the individual galaxies for a given stellar mass bin 
by size and summing their sSFRs until reaching 25\%, 50\% and 75\% of the total sSFR for that given stellar mass bin.  
This assumes that each galaxy within a bin is well represented by the median stacked sSFR, as the individual
galaxies adopt this median value.  
We see that 50\% of stars are formed in galaxies with sizes within $\pm$0.13 dex from the average 
star-forming size-mass relation, as shown as the solid blue from \citet{vanderWel14} for comparison.  
As the mean is a more physically relevant metric, we repeat the above analysis from the mean stacks of 
24$\mu$m finding a marginally wider spread of $\pm$0.14 dex.
This width of 0.26 dex (0.28 dex for the mean) is of similar order to the $1\sigma$ intrinsic
scatter of the size-mass relation \citep[see Figure 6c in][]{vanderWel14}.  Most of the stars
in the Universe are formed in star-forming galaxies with typical sizes.

While we find that amongst the overall galaxy population most of the star formation occurs within a narrow range
of sizes, we can further explore if we see any dependence on the star formation rate \emph{within} the star-forming population.  
In the following section, we isolate these star-forming galaxies based on their rest-frame U-V and V-J 
colors \citep[see Figure 26 in][]{Skelton14}.

\subsection{Do we see variations in star formation with galaxy size within the star-forming population alone?}
\label{sec:sfalone}

When solely selecting actively star-forming
galaxies based on their rest-frame colors (middle panels in Figure~\ref{fig:size}), 
the dependence of the median sSFR of a galaxy on size is far less pronounced 
relative to the overall galaxy population (left panels).
To first order, for any given stellar mass, larger star-forming galaxies have the
same sSFR as smaller star-forming galaxies within $<$0.2 dex.  
We start to see deviations from
this trend at the highest stellar masses and lowest redshift interval.
This is likely related to the observed flattening in the slope of the star formation sequence 
towards later times for galaxies more massive
than $\sim$10$^{10.5}$ M$_{\odot}$ \citep[e.g.,][]{Whitaker14b, Lee15}, 
and the correlation between this flattening and S\'{e}rsic index \citep{Whitaker15}.  We note that the
trends between the size-mass plane and sSFR for star-forming galaxies agree nicely with those 
at $z$=0 presented in Figure 14 of \citet{Omand14}. 

\begin{figure*}[t!]
\leavevmode
\centering
\includegraphics[width=0.8\linewidth]{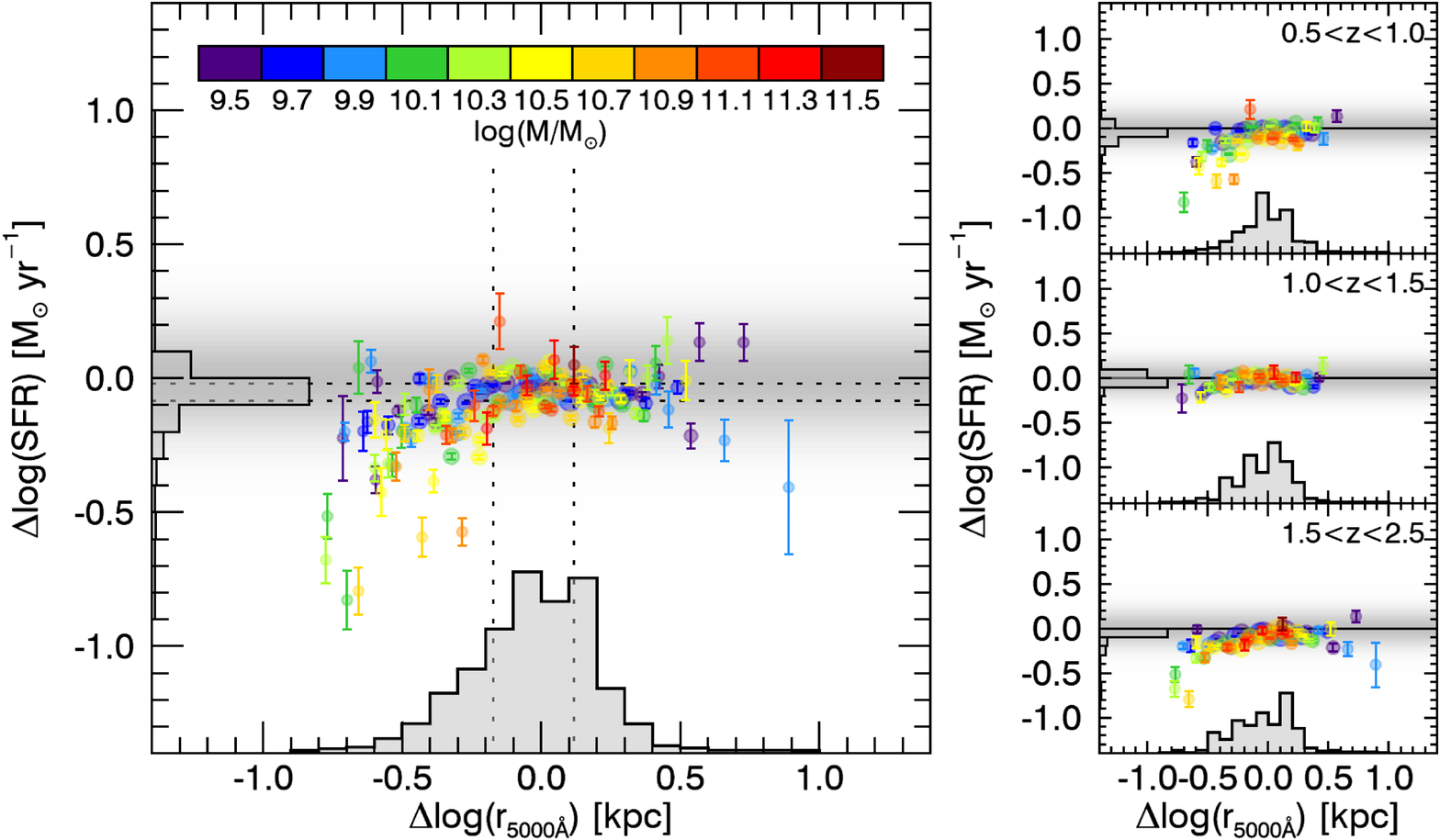}
\caption{The residual observed UV+IR star formation rate as a function of residual galaxy size from a stacking 
analysis across the size-mass plane in 0.2 dex bins for UVJ-selected star-forming galaxies.  Given the average
redshift and stellar mass of each measurement, the well-known correlations between log(SFR)-log(M$_{\star}$)
and log($r_{e}$)-log(M$_{\star}$) are subtracted to yield the residual values.  
Although the majority of galaxies show little dependence on 
galaxy size, we see that intermediate mass (log(M$_{\star}$/M$_{\odot}$)$\sim$10.0-10.6) compact galaxies
have star formation rates depressed by $>$0.5 dex below the average relations.  The left panel compiles
all redshifts from $z$=0.5 to $z$=2.5, whereas the right three panels break down the measurements by redshift bin. The
histogram shows the galaxy size distribution with the average relation subtracted. The greyscale
horizontal band indicates the observed 0.3 dex scatter in the star formation sequence for reference,
with a Gaussian transparency distribution, and
the dotted lines mark the 0.2 dex typical scatter in the observed size-mass relation.}
\label{fig:size_sfr}
\end{figure*}

\begin{figure}[t!]
\leavevmode
\centering
\includegraphics[width=\linewidth]{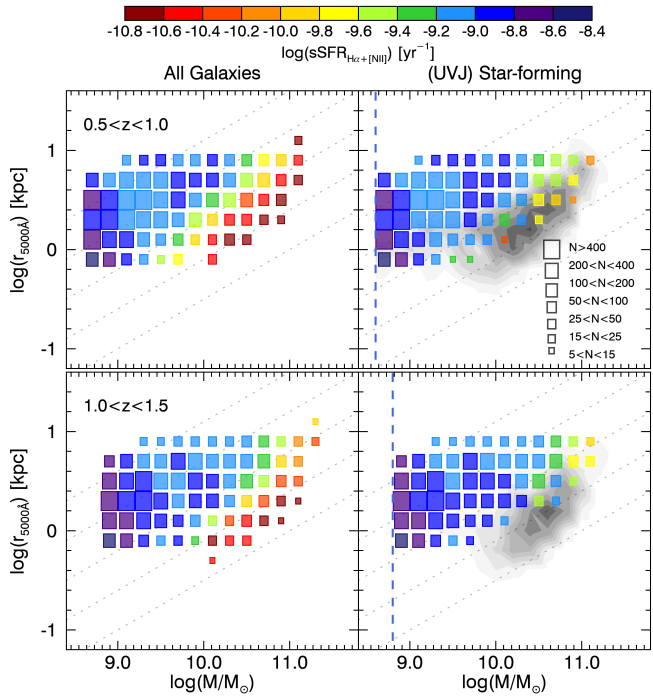}
\caption{When considering the dust-corrected H$\alpha$ sSFR, we find the same general trends as
the UV+IR SFR amongst the overall galaxy population: large galaxies have higher sSFRs than smaller galaxies
at fixed stellar mass. Similarly, we find the same lower envelope of compact star-forming galaxies
with depressed sSFRs.}
\label{fig:Halpha}
\end{figure}

In order to better quantify the dependence of the average star formation rate on galaxy size, we
first must remove the well-known trends with stellar mass and redshift.
As detailed in the Appendix, we reproduce the size-mass relation first presented in Figure~\ref{fig:size},
but instead color-code by $\Delta$log(SFR) (Figure~\ref{fig:size2}).  Next, we use the size-mass relation
of \citet{vanderWel14} to determine $\Delta$log($r_{e}$) for each 0.2 dex bin in stellar mass and size.
Figure~\ref{fig:size_sfr} presents this residual relation between the average star formation rate
and rest-frame 5000\AA\ galaxy size for star-forming galaxies only.  
The grey-scale demarcates the typical observed 0.3 dex scatter in the star-formation 
sequence \citep[e.g.,][]{Rodighiero11,Whitaker12b,Speagle14}, where the transparency is defined by a Gaussian 
distribution\footnote{The y-axis is measured from a stacking analysis, so we 
will not recover the intrinsic scatter in the SFR.}.
As each data point reflects one
of the original 0.2 dex bins in size and mass, we do not expect a scatter plot.  
But even with our stacking analysis, we would be sensitive to an overall correlation between 
scatter about the star formation sequence and scatter about the size-mass relation.

When taking into account the number of galaxies that go into each stack, we generate histograms of
the original data by adopting the median stacked SFR from each bin together with the input size and stellar mass 
distributions (shown in grey on the x- and y-axes).  
If we split the data into quartiles in $\Delta$log(SFR) (y-axis histogram), we find only a weak dependence on size; 
$\Delta$log($r_{e}$) for galaxies in the highest quartile are 0.27$\pm$0.06 dex larger than galaxies in the lowest quartile.
In other words, we see that the residual median SFRs of the majority of galaxies show little 
dependence on galaxy size (see also Fang et al. in prep).  If we instead split the quartiles based
on $\Delta$log($r_{e}$) (x-axis histogram), we similarly find a trend only amongst the smallest galaxies where 
$\Delta$log(SFR) is 0.11$\pm$0.02 dex lower that of the largest galaxy quartile.
This population of compact ``fading'' star-forming galaxies can also be clearly seen in 
the middle panels of Figure~\ref{fig:size2}.

Using spatially-resolved maps
of H$\alpha$ for star-forming galaxies at z$\sim$1 in the 3D-HST survey, \citet{Nelson15} 
find that H$\alpha$ is enhanced at all radii above the main ridgeline of the log(SFR)-log(M$_{\star}$)
plane, and depressed at all radii below.  This suggests that the physical processes driving the rate
of star formation acts throughout the entirety of galaxy disks.  Taking these results at face
value with the present analysis, this supports the idea that average galaxies follow ``parallel-tracks''
\citep[][see Section~\ref{sec:density} for further discussion]{vanDokkum15}.  
We note that this picture is complicated by the presence of large amounts of dust 
(and associated cold molecular gas) and apparent gradients 
thereof \citep[e.g.,][]{Tacconi10,Tacconi13,Simpson15,Tadaki15,Genzel15,Nelson16}, as well as potential systematic 
uncertainties introduced by the stacking methods employed. 
While results from \citet{Morishita15} demonstrate that the stellar mass density 
profiles do not appear to depend strongly on any potential color gradients, other studies have
found spatial variations in mass-to-light ratios to be important \citep[e.g.,][]{Wuyts12,Lang14}.  We
discuss this issue further in Section~\ref{sec:density}.

We have performed several tests in order to explore if the lack of trend between (s)SFR and 
galaxy size is driven by either the relatively large redshift bins or the adopted SFR indicator.
If we assume that the sSFR of any given galaxy is independent of the size, we can use the same parent
sample redshift distribution and the well observed average relation between log(SFR)--log(M$_{\star}$)
to randomly assign each galaxy a size from the \citet{vanderWel14} size-mass relation.  We find that
we can reproduce all observed trends, shy of the lower envelope of compact fading star-forming
galaxies.  The large redshift bins therefore do not effect the conclusions of this paper. 

The UV+IR total SFRs probe star formation timescales of order 100 Myr.  When instead considering the
H$\alpha$ star formation rate indicator, which is sensitive to shorter timescales of order 10 Myr,
we find the same trends amongst star-forming galaxies in Figure~\ref{fig:Halpha}.  
For this test, we use the H$\alpha$ emission line fluxes from the 3D-HST survey in the 
two lowest redshift bins and correct for dust attenuation from the best-fit A$_{V}$ from
the spectral energy distribution.  Even though this dust correction is somewhat uncertain 
\citep[see][]{Whitaker14b}, it is likely that we may be underestimating the dust attenuation which 
could make the trends between galaxy size and sSFR slightly more pronounced.  
However, from this test we can conclude that the general independence of star formation
rate and galaxy size is likely not sensitive to the timescale on which the star formation rate is probed.

The notable exception to the lack of dependence on sSFR are galaxies with intermediate
stellar masses (log(M$_{\star}$/M$_{\odot}$)$\sim$10.0-10.6) and compact sizes 
($r_{e}$$<$2 kpc). These massive compact galaxies have star formation 
rates significantly depressed relative to the bulk population of star-forming galaxies.  They
may be in the process of fading to join the quiescent population \citep[see also][]{Yano16}.
Larger statistical samples and/or deeper 24$\mu$m observations are required to further
improve the uncertainty in the star formation rates presented here.  Nonetheless, we
see evidence for an interesting trend between the star formation rates and sizes of intermediate mass, compact
star-forming galaxies.  

Now that we have confirmed the dichotomy between the sizes of quiescent and star-forming galaxies (Section~\ref{sec:size}),
with most of the star formation occurring within a narrow range of sizes (Section~\ref{sec:sizescale}), we 
additionally show that there is little dependence of sSFR on size within the star-forming galaxies alone.
Taking these three points together, this suggests an abrupt change in star formation rate after a galaxy 
attains a particular structure.  In the next section, we will investigate how well various galaxy 
structure parameters can uniquely predict this decrease in sSFR.

\section{Predicting Quiescence}
\label{sec:quiescence}

Amongst the several competing theories put forth to explain the simultaneous evolution of the structures of 
quiescent and star-forming galaxy populations across cosmic time, there are predictions in each case for
how galaxies are expected to grow in stellar mass with respect to their structures.
\citet{Barro15} postulate that the distribution of massive galaxies form an
``L-shaped track'' comprised of the two fundamental physical processes of compaction and quenching.
Galaxies will continue to gradually grow inside-out \citep[e.g.,][]{Nelson15}, until they
reach a strong phase of core growth.
Also known as compaction, this is a rapid period in which star-forming galaxies become structurally similar to quiescent
galaxies, growing in central
density (also increasing their core-to-total mass and S\'{e}rsic indices, and decreasing their size) 
\citep[see][]{Dekel14,Zolotov15}.
Quenching occurs when these compact star-forming galaxies reach a central density threshold.  

On the other hand, the ``parallel tracks'' model by \citet{vanDokkum15}\footnote{The likely progenitors of galaxies are predicted
in this model by tracing towards lower stellar masses and smaller sizes in the size-mass plane, hence following 
``parellel tracks'' for a given galaxy size at fixed stellar mass.} instead suggests that 
galaxies follow an inside-out growth track in the size-mass plane, where the stellar mass is 
gradually increased within a fixed physical radius and galaxies quench when they reach a 
stellar density or velocity dispersion threshold.  The relatively small observed scatter in 
the size-mass relation \citep{vanderWel14} and the star formation sequence \citep{Whitaker12b} support
this idea of a more gradual growth of galaxies.  

\begin{figure}[t!]
\leavevmode
\centering
\includegraphics[width=\linewidth]{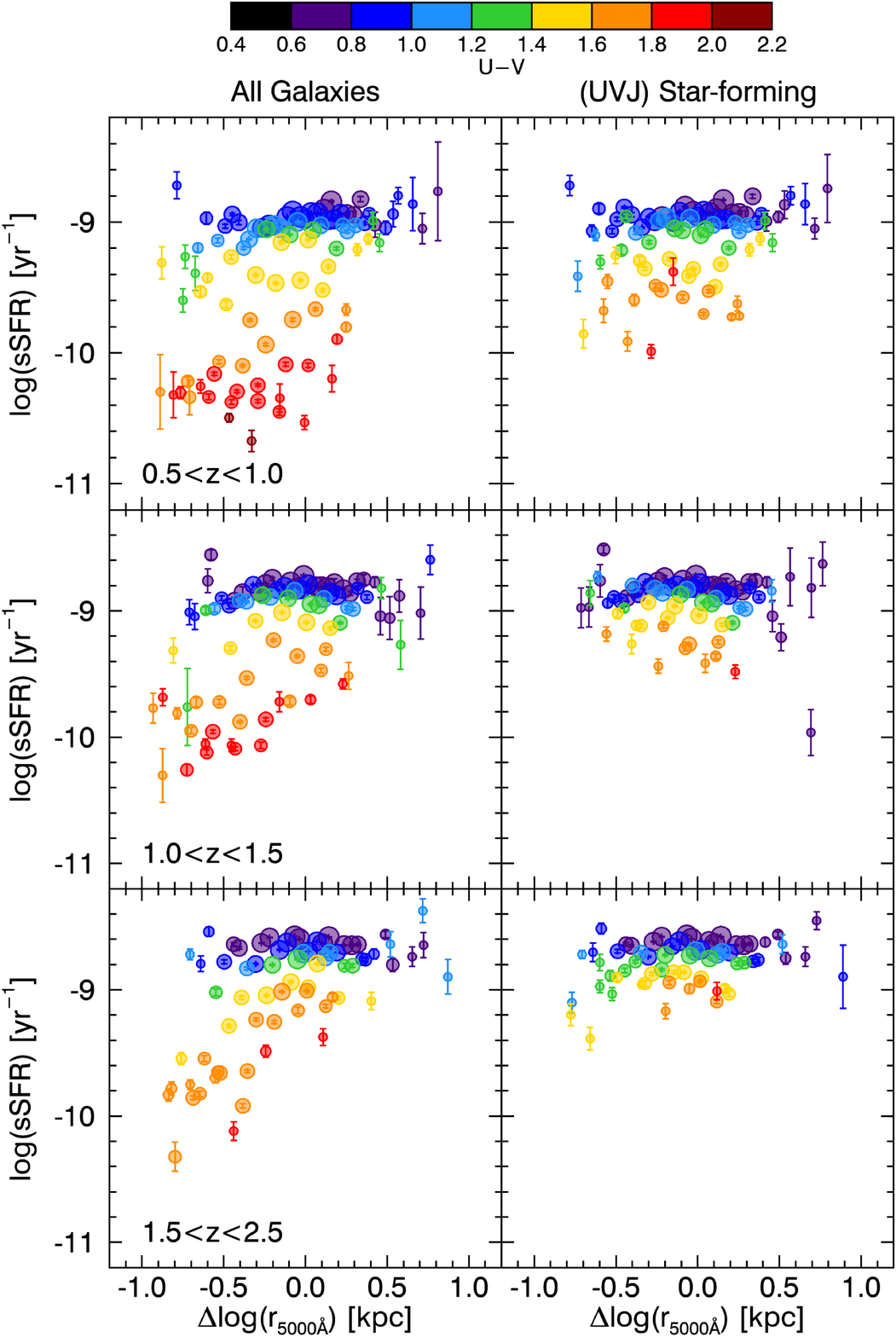}
\caption{The deviation of the logarithmic rest-frame 5000\AA galaxy size from the average size-mass
relation binned by 0.2 dex in mass and size and color-coded by rest-frame U-V color shows 
correlations with the median sSFR and rest-frame color. Negative values indicate 
galaxies that are more compact than average for their given stellar mass. The left panels show
all galaxies, whereas the right panels show UVJ-selected star-forming galaxies only.  The most compact
galaxies have the lowest sSFRs and reddest rest-frame colors, but there also exist similarly compact galaxies with
high sSFRs and blue rest-frame U-V colors.}
\label{fig:size_ssfr}
\end{figure}

\begin{figure}[t!]
\leavevmode
\centering
\includegraphics[width=\linewidth]{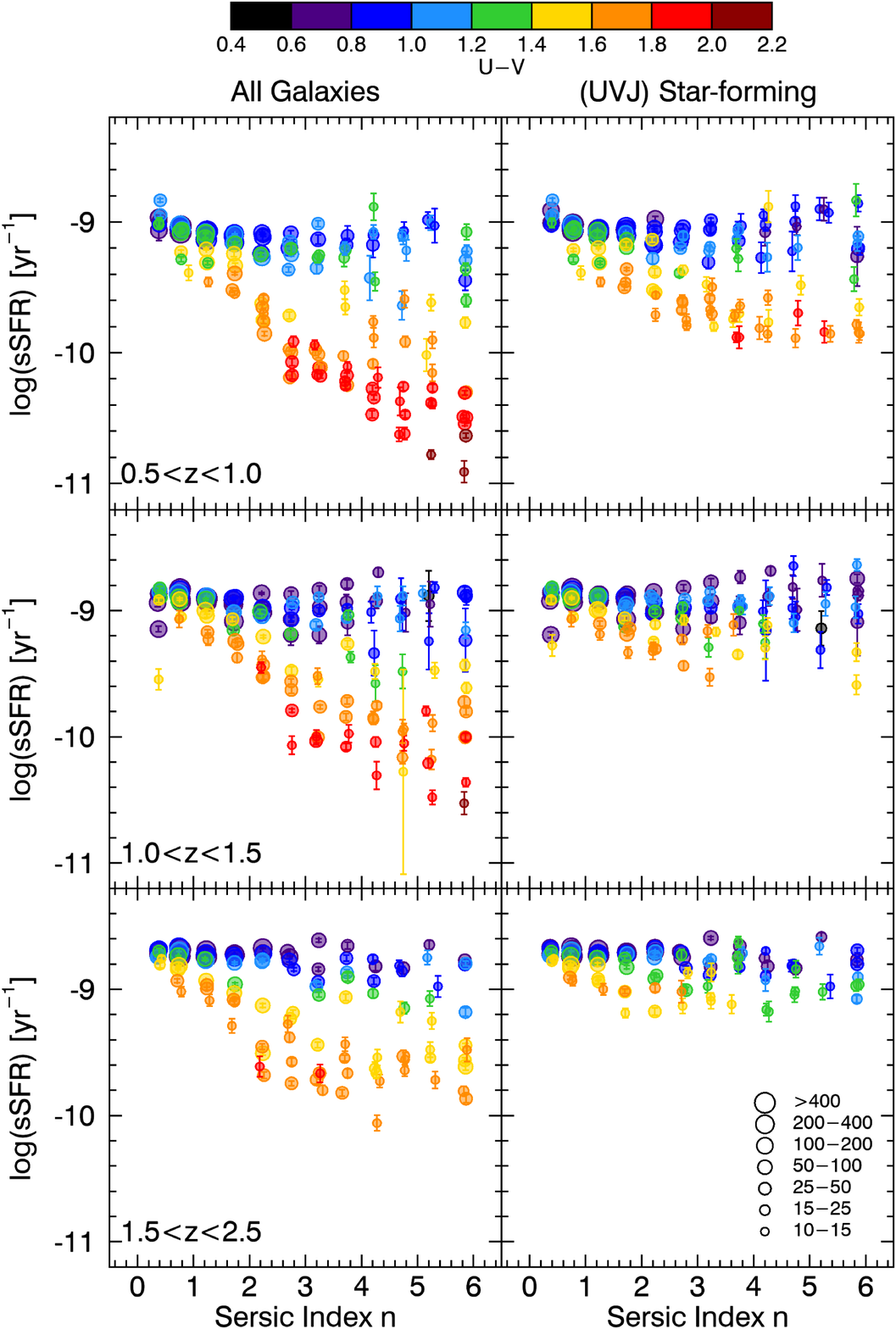}
\caption{The average S\'{e}rsic index $n$ for all galaxies (left) and UVJ-selected star-forming
galaxies (right) is well correlated with the median sSFR for a given rest-frame U-V colors 
across all redshifts probed here.  
Quantities are derived for galaxies stacked in the $n$-$\log$(M$_{\star}$) plane, 
with 0.2 dex bins for $\log$(M$_{\star}$) and 0.5 width bins in $n$.
As there is no universal trend, this implies that although $n$ is a good predictor of quiescence, it is not a
sufficient condition to predict a low sSFR.  The implications of these results on the slope and scatter of the 
star formation sequence are presented in \citet{Whitaker15}.}
\label{fig:sersic_ssfr}
\end{figure}

Regardless for the timescale defining the build up of the central mass concentration and the transition to quiescence, 
the common denominator between these various physical mechanisms is the existence of a threshold in either 
stellar density or velocity dispersion.  We will first explore if galaxy size or S\'{e}rsic 
index alone could predict quiescence in Section~\ref{sec:sersic}.  Next, we will perform a detailed 
analysis of the dependence of the stellar mass surface density and central density on sSFR in 
Section~\ref{sec:density}, parameterizing these correlations with broken power-law fits in 
Section~\ref{sec:evolution}.  Finally, we will define quiescence and the redshift evolution in the 
quenching threshold in Section~\ref{sec:evolve_quiescence}.

\subsection{What role do galaxy size and S\'{e}rsic index play in predicting quiescence?}
\label{sec:sersic}

From the present analysis, there are two interesting points regarding the ability to predict when a galaxy
will quench: (1) compact galaxies have lower sSFRs on average (Section~\ref{sec:size}), and (2) most 
stars form in galaxies with typical (larger) sizes for their given total stellar mass (Section~\ref{sec:sizescale}).
If a galaxy is observed to have a compact size for the given total stellar mass, does 
this uniquely predict it will be quenched?  We can recast the information in Figure~\ref{fig:size} to instead
plot the sSFR as a function of $\Delta$log($r_{e}$), or the deviation in size from the average 
size-mass relation (see Appendix for more details).  In Figure~\ref{fig:size_ssfr}, we see that galaxies with 
lower sSFRs are smaller than average with red rest-frame U-V colors.  However, it is important to note that 
there also exist a population of similarly compact galaxies with more typical sSFRs.  
This emphasizes that for a given stellar mass, galaxy size alone cannot predict quiescence.

Let us next consider S\'{e}rsic Index $n$:
in the case of ``morphological quenching'' \citep[e.g.,][]{Martig09, Martig13, Genzel14a},
a galaxy will quench once it builds up a significantly massive bulge, as indicated observationally by 
a high S\'{e}rsic Index $n$ \citep[e.g.,][]{Bruce14}.
As an earlier study by \citet{Whitaker15} did
not explicitly present the dependence of sSFR on $n$ in the same manner as the present analysis,
we therefore show this in Figure~\ref{fig:sersic_ssfr}, color-coded by the average rest-frame U-V color.
While high $n$ generally indicates lower sSFR, $n$ is 
also not a unique predictor; galaxies with high 
$n$ (bulge-dominated) exhibit a broad range of sSFRs that is strongly correlated with their rest-frame
U-V colors \citep[see also discussion in][]{Bell12}.  
The trends present in Figure~\ref{fig:sersic_ssfr} are striking, with important implications
for the derived slope and scatter of the star formation sequence.    
With this same data, \citet{Whitaker15} showed that the slope is of order unity for disk-like galaxies,
equivalent to a constant sSFR.  On the other hand, galaxies with $n$$>$2 (implying more dominant bulges and higher 
central densities) have significantly lower sSFRs than the main ridgeline of the star formation sequence. 
\citet{Brennan15} present a schematic diagram of how the various physical mechanisms could move galaxies around
the sSFR-$n$ plane, accounting for dry and wet mergers, disk instabilities, galaxy harassment, AGN feedback and other 
slow gas depletion processes.
However, the relevant take-away from the present data is that both galaxy size and $n$ alone are 
insufficient to isolate quiescent galaxies with low sSFRs.

\begin{figure}[t!]
\leavevmode
\centering
\includegraphics[width=0.95\linewidth]{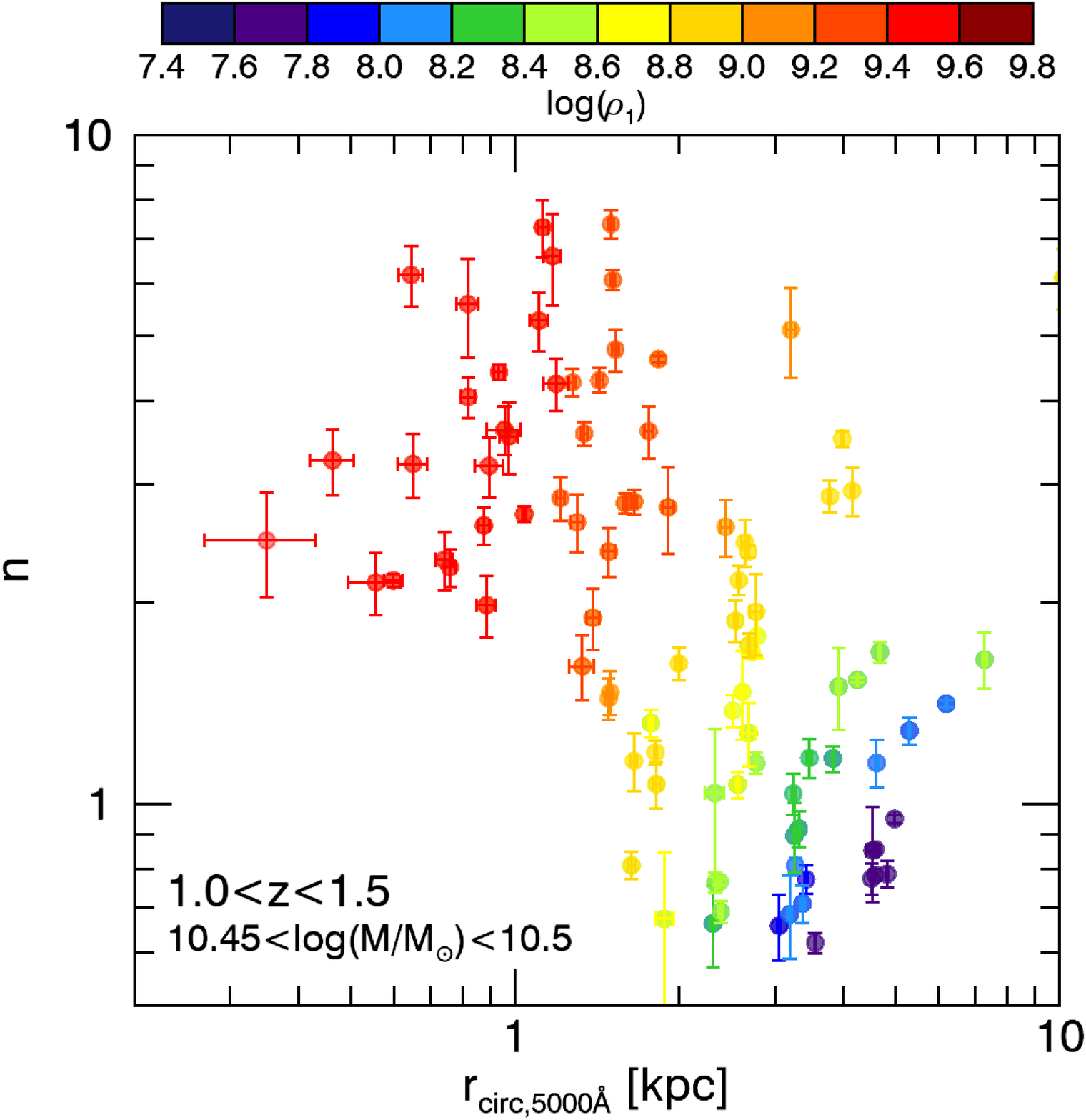}
\caption{The rest-frame 5000$\mathrm{\AA}$ size of a galaxy is strongly correlated with the
measured S\'{e}rsic index $n$, when selecting in a narrow range of stellar mass (selected to be 10.45$<$log(M/M$_{\odot}$)$<$10.5 for
demonstration purposes only).  The color-coding indicates the central density within a physical radius of 1 kpc for 
the individual galaxies: small galaxies with high $n$ have high central densities, whereas larger galaxies with low $n$ 
have lower central densities.}
\label{fig:demo}
\end{figure}

\begin{figure*}[t!]
\begin{minipage}{0.5\textwidth}
\centering
\includegraphics[width=\linewidth]{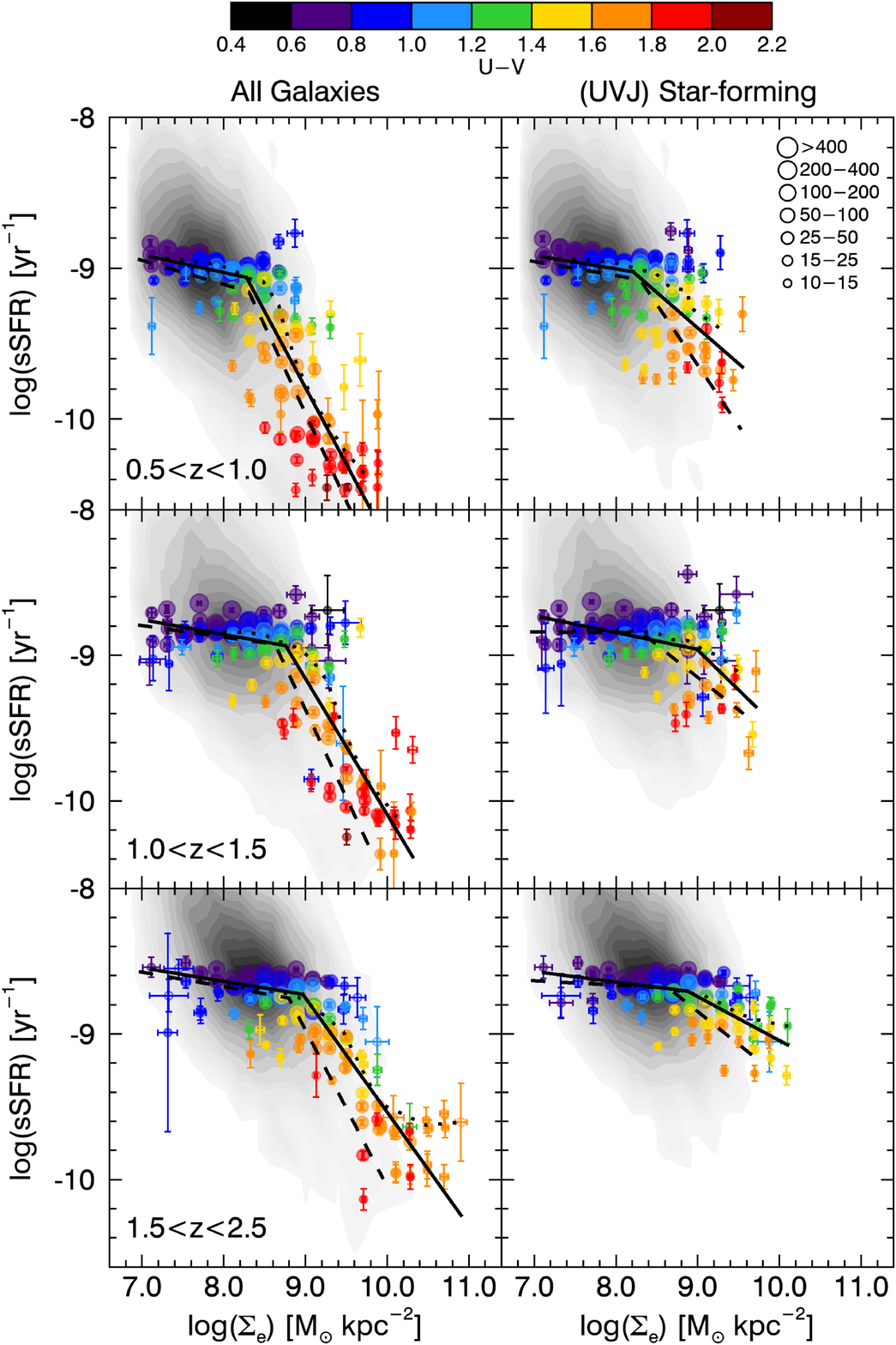}
\caption{We show the correlation between the median sSFR and surface density of all
galaxies (left) and UVJ-selected star-forming galaxies (right), color-coded by their average
rest-frame U-V colors.  sSFR is calculated here within 0.2 dex bins across the 
log($\Sigma_{e}$)-log(M$_{\star}$) plane, with the greyscale showing the contours for the overall
galaxy population with individual 24$\mu$m detections.
While there is considerable scatter, especially amongst the more compact
galaxies, there is a clear trend for a roughly constant sSFR for (blue) galaxies with surface densities
less than log($\Sigma_{e}$)$<$9 M$_{\odot}$ kpc$^{-2}$ and a drop off in sSFR for (red) galaxies with
higher surface densities. The dotted line is the running median.  The solid black line is a broken
power-law fit to the data, and the dashed line is the log(sSFR)-log($\Sigma_{1}$)
relation adapted from Figure~\ref{fig:sigma1_ssfr}, where log($\Sigma_{1}$)=log($\rho_{1}$)-log(4/3).}
\label{fig:sigmae_ssfr}
\end{minipage}
\begin{minipage}{0.5\textwidth}
\centering
\includegraphics[width=\linewidth]{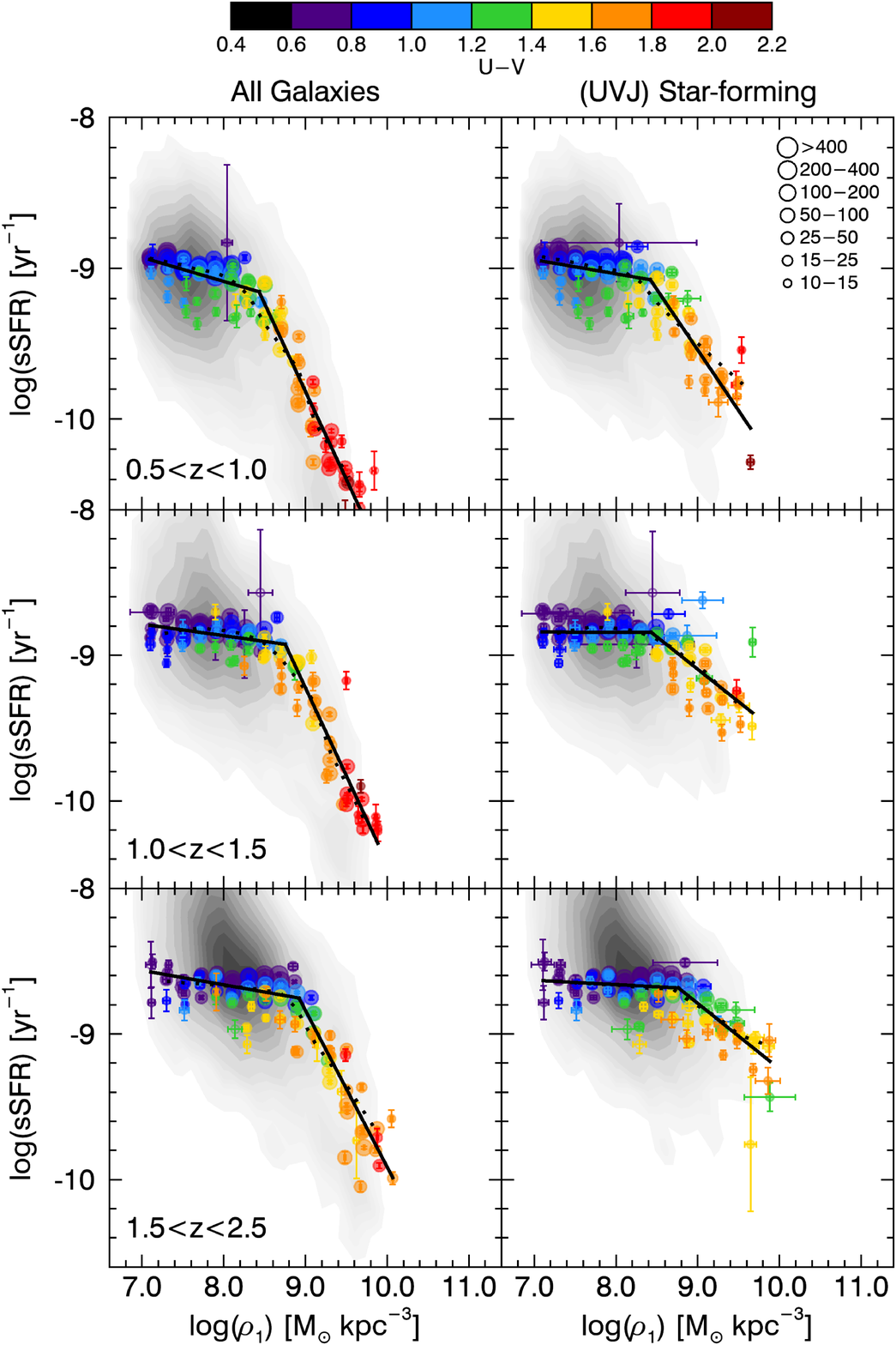}
\caption{Similar to Figure~\ref{fig:sigmae_ssfr}, we now show the correlation between the median sSFR
and central density, $\rho_{1}(r<1~\mathrm{kpc})$, for all galaxies (left) and 
UVJ-selected star-forming galaxies (right).  The data are split into three
redshift bins and color-coded by the average rest-frame U-V colors.
{sSFR is calculated here within 0.2 dex bins across the 
log($\rho_{1}$)-log(M$_{\star}$) plane.}
It is striking that the correlation between median sSFR and central density within 1 kpc is remarkably
tighter than that with surface density as determined from the effective radius in Figure~\ref{fig:sigmae_ssfr}.  
Although the trends are similar, the scatter amongst the measurements decreases significantly when considering
the central density of galaxies.}
\label{fig:sigma1_ssfr}
\end{minipage}
\end{figure*}

The stellar mass density profile of a galaxy is defined by both the effective radius and the
S\'{e}rsic Index $n$ of a galaxy.  The combination of these two parameters may therefore be more
powerful than either parameter alone.  Figure~\ref{fig:demo} demonstrates the strong correlation 
between galaxy size and S\'{e}rsic index at fixed stellar mass.  In this figure, we have selected 
a small number of galaxies from the parent sample such that they inhabit a narrow range in stellar mass
of 10.45$<$log(M/M$_{\odot}$)$<$10.50 at 1.0$<$$z$$<$1.5.  The galaxies are color-coded by their 
deprojected central density within a fixed physical radius of 1 kpc, as derived in Section~\ref{sec:centraldensity}. 
We see in Figure~\ref{fig:demo} that the stellar mass density within the central 1 kpc of these galaxies
is strongly correlated with both size and S\'{e}rsic index.
In the following sections, we will next explore if this stellar mass density is the most robust 
predictor of quiescence, relative to stellar mass, $r_{e}$ or $n$ alone. 

\subsection{Does a central density or circular velocity threshold uniquely predict quiescence?}
\label{sec:density}

We perform a similar analysis as in Figure~\ref{fig:size}, but instead derive median stacked UV+IR (s)SFRs 
in 0.2 dex bins of logarithmic stellar mass and stellar mass surface
density (log($\Sigma_{e}$); Figure~\ref{fig:sigmae_ssfr}) and central stellar density (log($\rho_{1}$); 
Figure~\ref{fig:sigma1_ssfr}) in three redshift intervals, 0.5$<$$z$$<$1.0, 1.0$<$$z$$<$1.5, and 1.5$<$$z$$<$2.5.
A similar bootstrap analysis on the stacked
IR SFRs was performed to derive uncertainties in the median values, as described in Section~\ref{sec:centraldensity}.  
These measured values are then recast as sSFRs as a function of log($\Sigma_{e}$) and 
log($\rho_{1}$) in the subsequent analysis.  By using
sSFR, we are normalizing the SFRs by stellar mass and probing an inverse timescale sensitive
to the ages of the stellar populations.  The individual measured values are shown in greyscale
in Figures~\ref{fig:sigmae_ssfr} and \ref{fig:sigma1_ssfr}; the plume rising towards higher sSFR
at the lowest densities, offset from the average relations, 
reflects the limits of the data on a galaxy-by-galaxy basis.  Similarly, we see few individual 
24$\mu$m detections at the lowest sSFRs and highest densities.  In both of these extreme parameter regimes, many of
the individual galaxies remain undetected in the 24$\mu$m imaging, necessitating the stacking analysis 
presented herein. 

Figure~\ref{fig:sigmae_ssfr} shows a strong correlation between the median sSFRs from galaxy stacks 
and their stellar mass density within the effective radius. 
The result is strikingly different to that of Figures~\ref{fig:size_ssfr} and \ref{fig:sersic_ssfr}. 
For galaxies with low surface densities, 
the sSFR is roughly constant at a given epoch.  On the other hand, we see a strong drop in the sSFR for galaxies
with high densities.  We observe these trends in all three redshift intervals, finding that
the turnover in sSFR at higher densities is stronger for quiescent and star-forming galaxies together than for star-forming
galaxies alone. As the majority of massive galaxies are quiescent \citep[e.g.,][]{Muzzin13} 
and have more centrally concentrated stellar 
mass profiles \citep[e.g.,][]{Bell12}, they will preferentially pull down the median sSFR at the highest densities. 
If we instead consider the central density, which is derived from the deprojected
stellar mass within a 1 kpc radius sphere, we see that the trend remains essentially the same shape but the 
scatter is significantly reduced in Figure~\ref{fig:sigma1_ssfr}.  The offset in stellar density when comparing
the solid and dashed black lines in Figure~\ref{fig:sigmae_ssfr}
result from the implicit difference between the stellar mass enclosed within the central 1 kpc as compared
to the effective radius.  We have already accounted here for
the constant factor of 0.12 dex introduced during the deprojection from a sphere to a circle.
While the trend in $\Delta$SFR-$\Delta$$\Sigma_{1}$ from \citet{Barro15} is coined
an ``L-track'', we show the more gradual evolution with remarkably small scatter
in sSFR with increasing central density when not correcting for stellar mass dependence, 
that we will also demonstrate to evolve with redshift.  
These results are in good quantitative agreement with B. Lee et al. (in prep), who perform a similar analysis but
instead use SFR indicators from SED fitting using flexible star formation histories.

\begin{figure}
\leavevmode
\centering
\includegraphics[width=\linewidth]{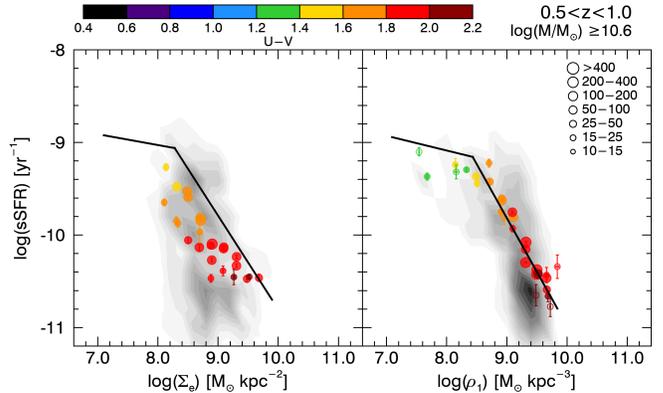}
\caption{This figure replicates the top left panel from Figure~\ref{fig:sigmae_ssfr} (left) and 
Figure~\ref{fig:sigma1_ssfr} (right), but only showing the stacks of massive galaxies with log(M/M$_{\odot}$)$\geq$10.6.
The data are grouped in 0.2 dex bins of stellar mass and log($\Sigma_{e}$) or log($\rho_{1}$), and
color coded by the average rest-frame U-V color.
While the most massive galaxies lie systematically below the
average log(sSFR)-log($\Sigma_{e}$) relation (left), we do not see any 
such offset when instead considering the central surface density within 1 kpc (right).   
The greyscale shows the contours for the massive galaxies with individual 24$\mu$m detections. 
Open symbols represent the 1$\sigma$ upper limit in the measured sSFR.}
\label{fig:sigma1_mass}
\end{figure}

Our results agree qualitatively with that of \citet{Woo15} at $z$=0, who show that SDSS central galaxies with
higher central surface densities have lower sSFR.
It is somewhat challenging to directly compare to the $z$=0
results with the present analysis, as they separate their sample by stellar mass.
In Figure~\ref{fig:sigma1_mass}, we isolate galaxies with log(M$_{\star}$/M$_{\odot}$)$\geq$10.6 
for both stellar surface mass density (left panel) and central surface density (right panel).  To 
facilitate a direct comparison, we convert central density to surface density here by applying an offset
of 0.12 dex corresponding to the equivalent difference in constants (see Equation~\ref{eq:rho1}).  While we only
show the one redshift epoch, 0.5$<$$z$$<$1.0, that has the largest dynamic range, we find 
similar trends at higher redshifts. 
We see that our measurements for the most massive galaxies span over two decades in stellar mass
central surface density and almost two decades in sSFR.  For similar stellar masses in \citet{Woo15}, the evolution
in central surface density at $z$=0 is significantly less, of order $<$0.5 dex.  
When considering where the most massive galaxies reside in the
log(sSFR)-log($\Sigma_{e}$) plane in the left panel of Figure~\ref{fig:sigma1_mass}, we see that they 
preferentially have lower sSFRs relative to 
the average relation, and thereby their less massive counterparts.   
Intriguingly, we no longer see this stellar mass dependence when instead 
considering central density (right panel, Figure~\ref{fig:sigma1_mass}).  

The earlier results of \citet{Lang14} showing that the bulge mass together with the bulge-to-total ratio (or $n$)
correlate most strongly with the degree of quiescence imply that the central mass concentration will
be a key factor in quenching.  Here we explicitly consider the density within the central 1 kpc out to $z$=2.5
and demonstrate that it is indeed a remarkably clean tracer of the median sSFR of galaxies. 
Although we show that $\rho_{1}$ tracks the median dependence of sSFR on structure best, it may be that
galaxies at a given $\rho_{1}$ exhibit a relatively broad range in intrinsic sSFRs. Given
that nature of the stacking analysis employed here, we cannot further study the intrinsic scatter in this relation.
Regardless, these results do confirm the low redshift study by \citet{Teimoorinia16}, who
use a novel technique to rank the relative importance of SDSS central galaxy properties in the process of 
quenching star formation. Similar to our conclusions, \citet{Teimoorinia16}  
find that central velocity dispersion and/or central stellar mass concentration are excellent 
predictors of the cessation of star formation. 

The present analysis rests on the assumption that light traces mass 
(Sections~\ref{sec:structure} and \ref{sec:centraldensity}). 
When accounting for any potential variations in mass-to-light ratios from either 
the star formation histories, ages, or dust across the galaxies,
\citet{Wuyts12} and \citet{Lang14} show that the stellar mass maps 
and profiles are generally smoother and more centrally concentrated 
than the profiles in rest-frame UV and optical light.  Furthermore, the evidence is mounting
for massive compact concentrations of gas and dust residing in a significant fraction of 
typical star-forming galaxies towards the high mass end from high-resolution studies of 
the rest-frame UV/optical light \citep{Guo11, Guo12, Wuyts12, Szomoru13, Tacchella15a, Boada15},
as well as (sub)mm studies \citep{Tacconi13, Simpson15, Tadaki15}.
The highest stellar mass star-forming galaxies may be effected most severely, where we could be underestimating their 
central mass concentration.
Preferentially underestimating the central densities and/or the SFRs \citep[e.g.,][]{Nelson16} 
of the most massive star-forming galaxies could make the turnover in sSFR above the density threshold (falsely) steeper. 
Figure~\ref{fig:sigma1_mass} demonstrates the dynamic range in central density that the most massive galaxies inhabit.   
Despite the effects of variations in mass-to-light ratios being stronger towards higher stellar masses, 
the location of the break in the log(sSFR)-log($\rho_{1}$) plane may not be affected severely as 
the vast majority of massive galaxies have central densities well above the turnover.

\subsection{Broken Power-law Fits}
\label{sec:evolution}

We fit the log(sSFR)--log($\Sigma_{e}$) relation in Figure~\ref{fig:sigmae_ssfr}
and log(sSFR)--log($\rho_{1}$) relation in Figure~\ref{fig:sigma1_ssfr}
with broken power laws to independently quantify the behavior of galaxies above
and below the characteristic values of $\Sigma_{e}$ and  $\rho_{1}$.  
The broken power law is parameterized as,

\begin{equation}
\log(\mathrm{sSFR}) = a(\log(\mathrm{X}) - b) + c 
\label{eq:powlaw}
\end{equation}

\noindent where X equals either $\Sigma_{e}$ or $\rho_{1}$. 
The slope $a$ is roughly flat below the characteristic densities $b$
(Figure~\ref{fig:bestfit_sigma_quenched}), and steeply drops
off above these characteristic values. Hereafter, these characteristic densities are 
notated as either $\Sigma_{e,\mathrm{char}}$ or $\rho_{1,\mathrm{char}}$.
The best-fit parameters and their associated uncertainties
are listed in Tables~\ref{tab:sigmae_powlaw} and \ref{tab:sigma1_powlaw}.
We see that the best-fit power low is very similar between the log(sSFR)--log($\rho_{1}$)
relation in Figure~\ref{fig:sigma1_ssfr}, and that of log(sSFR)--log($\Sigma_{e}$) relation in 
Figure~\ref{fig:sigmae_ssfr}. The best-fit log(sSFR)--log($\Sigma_{1}$) relation is shown as
a dashed line for reference in Figure~\ref{fig:sigmae_ssfr}, where log($\Sigma_{1}$)$\equiv$log($\rho_{1}$)-log(4/3).  
The main difference between
these two measurements is that there is a significantly larger scatter amongst the average measured values
of $\Sigma_{e}$, whereas $\rho_{1}$ (or $\Sigma_{1}$) shows remarkably little scatter.  

\begin{table*}[t]
\centering
\begin{threeparttable}
    \caption{Broken Power Law Fits: log(sSFR)-log($\Sigma_{e}$)}\label{tab:sigmae_powlaw}
    \begin{tabular}{lcccccccc}
      \hline \hline
      & \multicolumn{4}{c}{All Galaxies} &  \multicolumn{4}{c}{UVJ Star-Forming} \\
      \cline{3-4} \cline{7-8}
      redshift range  & $\mathrm{a_{low}}$ & $\mathrm{a_{high}}$ & $b$  & $c$ & $\mathrm{a_{low}}$ & $\mathrm{a_{high}}$ & $b$  & $c$\\
      \hline
      \noalign{\smallskip}
      $0.5<z<1.0$  & $-0.12\pm0.04$ & $-1.01\pm0.03$ & $8.28\pm0.03$ & $-9.06\pm0.02$ & $-0.09\pm0.04$ & $-0.47\pm0.04$ & $8.21\pm0.07$ & $-9.02\pm0.02$\\
      $1.0<z<1.5$  & $-0.10\pm0.03$ & $-0.93\pm0.05$ & $8.76\pm0.04$ & $-8.93\pm0.02$ & $-0.12\pm0.02$ & $-0.55\pm0.17$ & $8.99\pm0.12$ & $-8.96\pm0.03$\\
      $1.5<z<2.5$  & $-0.09\pm0.03$ & $-0.78\pm0.04$ & $8.97\pm0.04$ & $-8.73\pm0.02$ & $-0.07\pm0.03$ & $-0.30\pm0.05$ & $8.87\pm0.11$ & $-8.71\pm0.02$\\
      \noalign{\smallskip}
      \hline
      \noalign{\smallskip}
    \end{tabular}
    \begin{tablenotes}
      \small
    \item \emph{Notes.} Broken power law coefficients parameterizing the evolution of the $\log$(sSFR)-$\log(\Sigma_{e})$
      relation from the median stacking analysis (Equation~\ref{eq:powlaw}).  
      $\mathrm{a_{low}}$ signifies the best-fit slope for galaxies 
      below the characteristic stellar surface mass density $\Sigma_{e}$, and $\mathrm{a_{high}}$ is the slope
      above this limit.
    \end{tablenotes}
  \end{threeparttable}
\end{table*}

\begin{table*}[t]
\centering
\begin{threeparttable}
    \caption{Broken Power Law Fits: log(sSFR)-log($\rho_{1}$)}\label{tab:sigma1_powlaw}
    \begin{tabular}{lcccccccc}
      \hline \hline
      & \multicolumn{4}{c}{All Galaxies} &  \multicolumn{4}{c}{UVJ Star-Forming} \\
      \cline{3-4} \cline{7-8}
      redshift range  & $\mathrm{a_{low}}$ & $\mathrm{a_{high}}$ & $b$  & $c$ & $\mathrm{a_{low}}$ & $\mathrm{a_{high}}$ & $b$  & $c$\\
      \hline
      \noalign{\smallskip}
      $0.5<z<1.0$  & $-0.16\pm0.03$ & $-1.16\pm0.04$ & $8.43\pm0.04$ & $-9.16\pm0.03$ & $-0.09\pm0.03$ & $-0.80\pm0.05$ & $8.42\pm0.05$ & $-9.08\pm0.02$\\
      $1.0<z<1.5$  & $-0.08\pm0.03$ & $-1.21\pm0.06$ & $8.75\pm0.04$ & $-8.93\pm0.02$ & $-0.00\pm0.04$ & $-0.46\pm0.05$ & $8.43\pm0.08$ & $-8.84\pm0.02$\\
      $1.5<z<2.5$  & $-0.10\pm0.02$ & $-1.08\pm0.07$ & $8.92\pm0.05$ & $-8.75\pm0.02$ & $-0.03\pm0.03$ & $-0.45\pm0.06$ & $8.77\pm0.09$ & $-8.68\pm0.02$\\
      \noalign{\smallskip}
      \hline
      \noalign{\smallskip}
    \end{tabular}
    \begin{tablenotes}
      \small
    \item \emph{Notes.} Broken power law coefficients parameterizing the evolution of the $\log$(sSFR)-$\log(\rho_{1})$
      relation from the median stacking analysis (Equation~\ref{eq:powlaw}).  
      $\mathrm{a_{low}}$ signifies the best-fit slope for galaxies 
      below the characteristic stellar mass density $\Sigma_{1}$, and $\mathrm{a_{high}}$ is the slope
      above this limit.
    \end{tablenotes}
  \end{threeparttable}
\end{table*}

In order to test that the trends we observe in Figures~\ref{fig:sigma1_ssfr} 
are not driven by flat galaxies, where the deprojected central density may 
not be robust given the assumptions, we repeat the analysis removing all galaxies with $n$$<$2.  
We find that the best-fit power laws for the entire sample are unchanged, albeit we are missing
the vast majority of data points falling below log($\rho_{1}$)$<$7.5 M$_{\odot}$ kpc$^{-3}$ where
most galaxies are disks with low $n$.  
Furthermore, as the numerical integration to measure $\rho_{1}$ depends not only on 
the reliability of $r_{e}$ but also $n$, we must test a more conservative stellar mass limit.  
If we remove all galaxies within 0.5 dex of the stellar mass limits, as required to ensure sufficient
S/N for robust measurements of $n$ for all galaxies, we find the best-fit power law fit to be statistically
unchanged.  We therefore assert that the observed correlations in this analysis appear stable against 
systematic biases in the structural parameters.

\subsection{Do we see redshift evolution of the quenching threshold in central density and circular velocity?}
\label{sec:evolve_quiescence}

While the characteristic central density, $\rho_{1,\mathrm{char}}$, marks 
a threshold above which galaxies become less efficient at forming new stars, it does not necessarily signal quiescence.
One could simply define a galaxy to have reached quiescence when log(sSFR)$<$-10 yr$^{-1}$. 
Alternatively, a galaxy could be defined as quenched once
it deviates significantly below the average star formation sequence at that cosmic epoch.  This 
latter definition will result in an evolving limit in log(sSFR) with cosmic time.
To that end, here we estimate the quenching limit in sSFR in one of two ways: (1) at a fixed limit of log(sSFR)$<$-10 yr$^{-1}$, and (2)
for an evolving limit in log(sSFR) based on results presented in \citet{Whitaker14b}.
The critical central density for quenching, which we define to be $\rho_{1,\mathrm{quenched}}$,
is then taken to be the central density at these respective sSFR limits in Figure~\ref{fig:sigma1_ssfr}.
For the purpose of this exercise, we only consider lower mass galaxies where the sSFR is roughly 
constant when calculating the evolving limit, ignoring the strong evolution in the turnover in sSFR at the massive end.  
If we assume the average log(sSFR) equals -9.0, -8.8, and -8.6 yr$^{-1}$ at 0.5$<$$z$$<$1.0, 1.0$<$$z$$<$1.5, and
1.5$<$$z$$<$2.5, respectively, an offset of 1 dex below the star formation sequence ($\sim$3$\sigma$) reaches
log(sSFR)=-9.6 yr$^{-1}$ at $z$=2 and lower values at later times (log(sSFR)=-10 yr$^{-1}$ by $z$=0.75).  
\citet{Barro15} adopt a 0.7 dex offset below the star formation sequence, however they normalize at the 
massive end where there is an evolving turnover towards increasingly lower sSFRs relative to the 
lower mass population as redshift decreases.

\begin{figure}[t]
\leavevmode
\centering
\includegraphics[width=\linewidth]{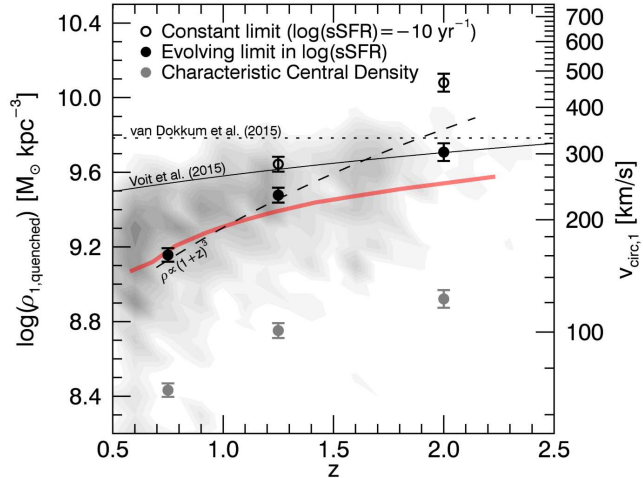}
\caption{We find strong evolution in the central density threshold for quenching.  
If we define a galaxy as quenched when log(sSFR)$<$-10 yr$^{-1}$, the
corresponding central density measured from Figure~\ref{fig:sigma1_ssfr}, 
$\log(\rho_{1,\mathrm{quenched}})$, is significantly 
higher at earlier times (open circles).  We find a similar trend if we instead 
define quenching to occur when galaxies
have sSFRs 1 dex below the average observed sSFR for low-mass galaxies
(filled circles; overlapping by definition at $z$=0.75).  
The grey filled circles indicate the characteristic density, $\log(\rho_{1,\mathrm{char}})$, 
marking the turnover in the broken power law fit.
The greyscale shows the central
densities measured for quiescent galaxies at analogous epochs, where the red line marks the running 
median of the quiescent population; the median central density of the quiescent population lies just below the quenching
threshold at $z$$>$1. The dashed line
indicates the expected cosmological evolution in density normalized at $z$=0.5, where $\rho\propto(1+z)^{3}$.
We also show the equivalent central circular velocity on the right axis, where $v_{\mathrm{circ},1}\propto\sqrt{\rho_{1}}$.
The dotted line is the assumed constant threshold in 
velocity dispersion above which galaxies at 1.5$<$$z$$<$3.0 quench from \citet{vanDokkum15}. 
The thin black line is the predicted
quenching threshold from \citet{Voit15}, normalized to 300 km/s at $z$=2.}
\label{fig:bestfit_sigma_quenched}
\end{figure}

Regardless of the adopted definition of quiescence, the average trends in 
Figure~\ref{fig:bestfit_sigma_quenched} indicate that the quenching threshold is almost a decade 
higher at $z$$\sim$2 compared to $z$$\sim$0.7. 
This implies that, on average, quiescent galaxies that quench at later times will have lower central
densities and velocities.  Furthermore, the quenching density as defined by a fixed offset from the 
average star formation sequence (and correspondingly evolving sSFR limit) has the same redshift
evolution as the characteristic density (grey points in Figure~\ref{fig:bestfit_sigma_quenched}).

We further compare our results to density and velocity thresholds presented in the literature.
After factoring out the stellar mass dependence of $\log(\Sigma_{1})$, 
\citet{Barro15} assert that galaxies quench once they reach a central surface density of 
$\log(\Sigma_{1})$$>$9.5 M$_{\odot}$ kpc$^{-2}$ (or $\log(\rho_{1})$$>$9.38 M$_{\odot}$ kpc$^{-3}$, when 
accounting for the deprojection of a sphere of radius 1 kpc).  
It is this correction for the stellar mass dependence of $\log(\Sigma_{1})$ that accounts
for the difference between the more gradual turnover in sSFR above the characteristic central density we observe 
here, and the abrupt turnover in $\Delta\log(\Sigma_{1})$ presented in Figure 7 of \citet{Barro15}. 
Perhaps unsurprisingly, the trends we observe in Figure~\ref{fig:bestfit_sigma_quenched} 
are close to the expected cosmological evolution of density, 
as $\rho$$\propto$$(1+z)^{3}$ (dashed line, normalized to the lowest redshift measurement).  
Similarly, as $v_{\mathrm{circ},1}$$\propto$$\sqrt{\rho}$,
the central circular velocity scales as $(1+z)^{3/2}$.  As the central 
circular velocity is directly proportional to the central density (see Section~\ref{sec:vcirc}), 
we show the corresponding values of $v_{\mathrm{circ},1}$ in the right axis of Figure~\ref{fig:bestfit_sigma_quenched}.
\citet{vanDokkum15} quote a threshold in velocity dispersion of 225 km/s, based on
an analysis of compact star-forming galaxies at 1.5$<$$z$$<$3.0
with a median size is $r_{e}$ = 1.8 kpc.  When correcting
to $r$ = 1 kpc following \citet{Cappellari06}, this quenching threshold increases to 234 km/s.  This 
is equivalent to a central circular velocity of $\sim$331 km/s, assuming $v_{\mathrm{circ}}$=$\sqrt{2}$$\sigma$.

\begin{figure}[t]
\leavevmode
\centering
\includegraphics[width=0.95\linewidth]{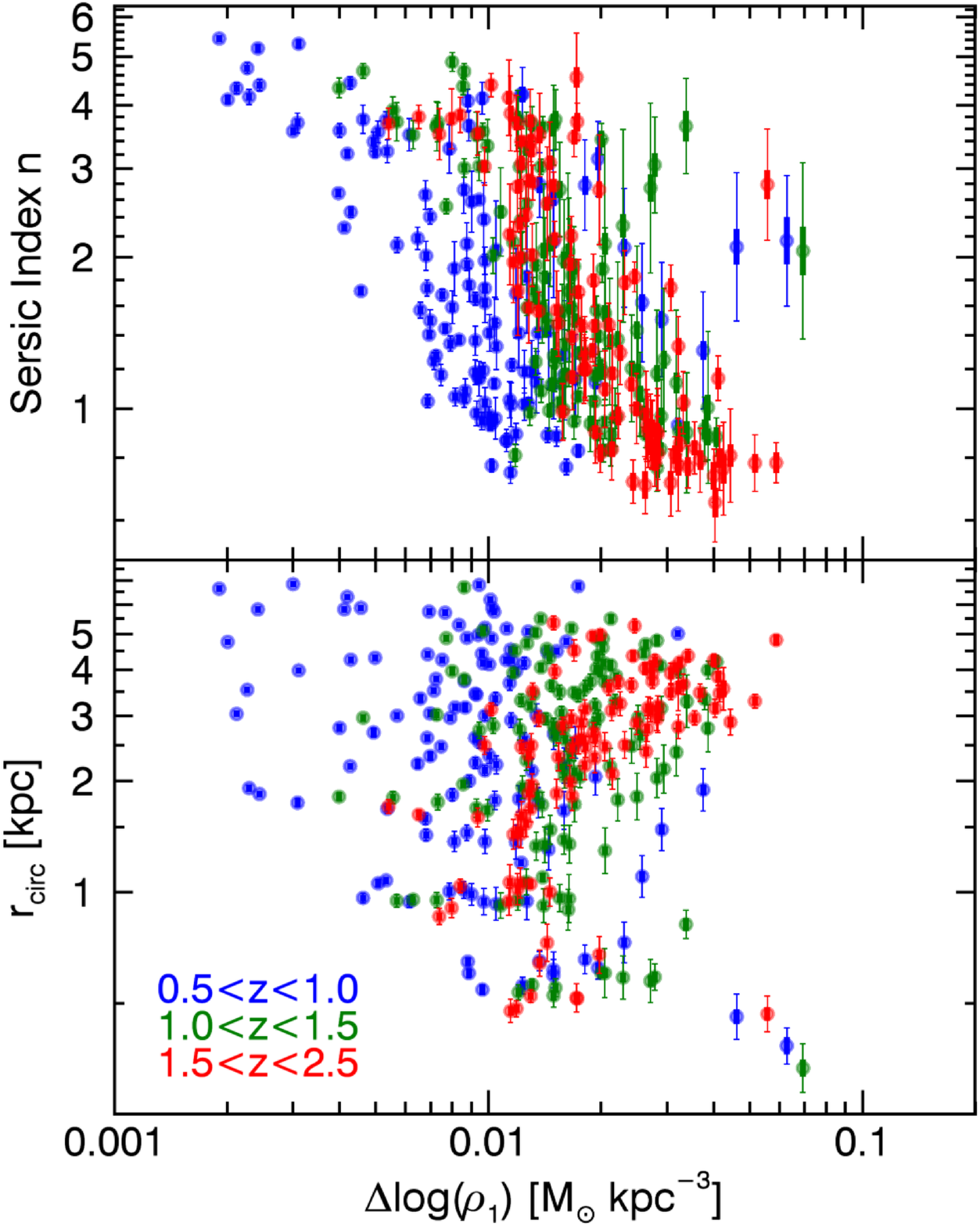}
\caption{The error analysis on the data presented in Figure~\ref{fig:sigma1_ssfr} reveals that
galaxies with the lowest $n$ generally have the largest uncertainty in $\rho_{1}$.  The thin error bars 
represent the errors of all galaxies in that bin added in quadrature, whereas the thick error bars represent
the error in the mean.  The trend between
the uncertainty in $\rho_{1}$ and circularized $r_{e}$ is less clear, aside from the most extreme small
galaxy bins having the largest errors in $\rho_{1}$. The error in $\rho_{1}$ is generally larger at $z$$>$1.}
\label{fig:error}
\end{figure}

Finally, we show the central densities of quiescent galaxies in grayscale in 
Figure~\ref{fig:bestfit_sigma_quenched}, with the running median shown in red.  
Although we believe that these high central densities are robust, as the measured $r_{e}$ and $n$ were 
only considered in the case of a sufficiently high S/N, we cannot rule out that they are biased 
low at the highest redshifts due to resolution limitations.  However, Figure~\ref{fig:error} shows the
results of the error analysis on the data presented in Figure~\ref{fig:sigma1_ssfr}, showing that the error of 
the mean $\rho_{1}$ is smallest for galaxies with high $n$.  The error analysis accounts for the covariance
of the parameters, as described in Section~\ref{sec:centraldensity}.  Even with the bin of the smallest 
galaxies having the largest uncertainty in $\rho_{1}$, the errors are not large enough 
to significantly effect the trends we see in Figure~\ref{fig:bestfit_sigma_quenched}.
The median central density of quiescent galaxies at a given epoch is similar to our 
quenching threshold based on an evolving threshold in sSFR.  
If we instead saw that
these quiescent galaxies had even higher central densities, this would suggest that the threshold
we find is too low, as otherwise the star-forming galaxies would need to be able to continue to form more stars
and further increase their central mass concentrations. 
The results of \citet{Bezanson09} show that the central densities of high redshift
galaxies are slightly higher than low redshift ellipticals.  As we observe galaxies to puff up over time 
via minor mergers and accretion \citep[e.g.,][]{Newman12}, this suggests that although quiescent 
galaxies grow in size with time, 
their central densities will likely not continue to increase once quenched, and may decrease slightly.
As these quiescent galaxies must have reached this threshold
at earlier times, their distribution of central densities with redshift suggest that the 
evolution of the quenching threshold 
slows down above $z$$\sim$2.  It is unclear if this evolution in the quenching threshold
is physical or simply an artifact of other processes triggering the shutdown of star formation.
Despite the fact that is difficult to interpret the meaning behind this evolving quenching threshold, 
the observed central densities of quiescent galaxies with redshift hint that quiescence
should not be defined as a non-evolving limit of log(sSFR)=-10.
Spatially resolved absorption line studies of the steller populations with extremely deep datasets such 
as the upcoming LEGA-C survey at $z$$\sim$1 \citep{vanderWel16} will potentially constrain the timescale for 
quenching as well as to better quantify the physical parameters predicting quiescence.

\section{Discussion}
\label{sec:discussion}
 
Theoretical predictions for the interplay between galaxy structures and their star formation histories are far from 
reaching a consensus.  In this section we summarize some of the key predictions and compare them to the empirical 
results from this paper. We place emphasis on the quenching process, which must both truncate star formation and 
structurally transform galaxies as they migrate from a star-forming population to a quiescent one.  Although the 
current analysis does not suggest an overall residual correlation between SFR and galaxy effective radius,
we have identified a population of compact, intermediate-mass star-forming galaxies with depressed SFRs. 
Focusing specifically on this population of compact, likely quenching galaxies, we discuss 
whether theoretical studies predict their existence.

There are two main channels in cosmological simulations to form massive compact galaxy populations: 
(1) the galaxies have very early formation times when the Universe 
was far denser \citep{Khochfar06, Wellons15}, or (2) they are the result of a central starburst 
driven by violent disk instabilities \citep{Zolotov15,Ceverino15} or gas-rich mergers \citep{Wellons15}. 
However, it may also be that galaxies do not undergo such ``compaction'' events, and that compact galaxies 
simply evolved from lower mass, slightly smaller galaxies \citep{vanDokkum15}.

In cosmological simulations by \citet{Tacchella16}, where a central starburst drives structural evolution,
it is the gas and young stars in galaxies with high sSFRs (above the average star formation sequence) 
that are predicted to be compact with short gas depletion timescales.
\citet{Tacchella16} do not however find any gradients in the stellar mass distribution, tracing the older stellar
distribution.  If we consider only the galaxies with compact rest-frame 5000\AA\ sizes in this study, we
similarly do not see evidence that they have higher than average sSFRs.
If anything, we see the opposite 
trend, at least amongst the most compact intermediate stellar mass galaxies 
(log(M$_{\star}$/M$_{\odot}$)$\sim$10.0-10.6).  Such fading galaxies
appear to not be present in the Tacchella simulations, based on their mass-weighted sizes.
Results from the EAGLE simulation (at $z$=0), on the other hand, predict a stronger dependence of galaxy size on
sSFR than our higher redshift observations \citep[Figure 2 in][]{Furlong15},
with $\Delta$log(sSFR)/$\Delta$log($r_{e}$) at fixed 
stellar mass ranging between $\sim$0.6--1.4 compared to typical values of $\sim$0.1--0.5 in the observations.
Using semi-analytic models, \citet{Brennan17} show 
a weak(er) trend amongst the most compact
galaxies at $z$=0--2.5 that falls between these two extremes, though the comparison cannot be made directly as the 
stellar mass dependence has not been factored out.  In summary, theoretical results predict a 
range from no residual dependence of galaxy size on SFR to moderately strong trends.

One key trend that the EAGLE simulations do not reproduce amongst the star-forming population is 
the lack of variation in sSFR at lower stellar masses.  
While this model shows variations of order 0.3--0.5 dex, the variation in sSFR within the five 
extragalactic fields included in the 3D-HST dataset is $<$0.2 dex.  
Unfortunately, \citet{Furlong15} do not present their higher redshift results, so a more direct 
comparison at the equivalent epochs is not possible.  Similarly, this information cannot be 
reconstructed from the results of \citet{Tacchella16} and \citet{Brennan17}.
Future such comparisons between the observations and theoretical models will prove illuminating.

Returning to the issue of gas depletion in relation to compaction, 
a recent study by \citet{Spilker16} find extremely low (CO) gas fractions in a pilot 
sample of compact star-forming galaxies, suggesting short gas depletion timescales.
As these compact star-forming galaxies exist in very small numbers, they would need to quench rapidly 
($<$0.5 Gyr timescales) in order to produce the required number of compact quiescent galaxies \citep{vanDokkum15}.
Indeed, the early results on the gas depletion timescales suggest timescales of order 100 Myr or 
less \citep{Spilker16, Barro16}. 
\citet{Saintonge12} also show that galaxies undergoing mergers or showing signs of morphological disruptions have 
the shortest molecular gas depletion times.
These results hint that the timescale for galaxies to pass through this compact high sSFR phase is short, 
and this is why the observational evidence is lacking when considering the average trends presented herein.

Next, we turn our focus back to the full galaxy population.
Both the remarkably small scatter and the evolution of the average relation between sSFR and central density is 
interesting in the context of recent arguments in the literature regarding the 
the nature of the most recently quenched galaxies and their role in the evolution of the size-mass 
relation of quiescent galaxies.  Although it has been shown that quiescent galaxies will experience 
growth through minor mergers and accretion \citep{Bezanson09, Newman12}, the simplest 
explanation of their size growth is the continuous addition of (larger) recently quenched 
galaxies \citep{vanderWel09a}. Galaxies that quench at later times are expected to have 
larger sizes because the Universe was less dense and therefore gas-rich, dissipative 
processes were less efficient \citep{Khochfar06}.  Indeed, observations at $z$$<$1 find that 
the most recently quenched galaxies are the largest \citep{Carollo13}.  There is a mixed 
bag of size measurements at $z$$\sim$1.5 \citep{Belli15}, with recently quenched galaxies 
exhibiting a range of sizes, and results at $z$$>$1.5 find that the most recently quenched 
galaxies are similar, if not more compact, than older quiescent galaxies at the same 
epoch \citep{Whitaker12b, Yano16}. 
\citet{Furlong15} further predict a trend (at $z$=0) for higher sSFR (suggesting more recent assembly) for 
larger quiescent galaxies at fixed stellar mass.  
We do not see any strong trends amongst our quiescent observations at $z$$>$0.5.
Although, as the 24$\mu$m derived SFRs are likely upper limits for quiescent galaxies, it is possible 
that we are washing out a stronger intrinsic trend within the observations.
It may be that recently quenched galaxies are more compact at high redshift, whereas they 
are larger at later times.
At $z$$>$1, we show here that the central density threshold for quenching is 
higher relative to the already quenched population in Figure~\ref{fig:bestfit_sigma_quenched} 
(shown as greyscale, with the running median in red).
We therefore see evidence that higher redshift quiescent galaxies are predicted to have 
significantly higher central densities to quench, which may therefore alleviate some of the tension 
in the observations between low and higher redshift analyses.  

When considering where massive galaxies populate the $\log$(sSFR)--$\log$($\rho_{1}$) plane 
(e.g., Figure~\ref{fig:sigma1_mass}), we see that they tend to have the densest central concentrations of
stellar mass.  It is important to note however, that the trend itself between central density and 
sSFR does not vary strongly with stellar mass at a given epoch.  In other words, while different stellar
mass regimes tend to populate the upper and lower end of this relation, it is the entire relation
itself that evolves with redshift.  Although there is no straight
forward way to plot the results of \citet{Barro15} in the $\Delta\log$(SFR)--$\Delta\log$($\Sigma_{1}$) plane
in our Figure~\ref{fig:bestfit_sigma_quenched}, they are in good qualitative agreement given the definition
of $\Delta\log$($\Sigma_{1}$).  
The stellar mass dependence of this quenching
threshold in central density observed by \citet{Fang13} at $z$=0 may also be explained in 
part by more massive galaxies having formed earlier in the Universe \citep{Kauffmann03}.
There is a thorough discussion of this downsizing in quenching in \citet{Dekel14}. 
As more massive dark matter halos will tend to cross a threshold halo mass for viral shock heating
earlier in the Universe, halo quenching will occur preferentially
in more massive galaxies \citep[e.g.,][]{Neistein06, Bouche10}.  Similarly, violent disk instabilities
will also have a natural downsizing, which \citet{Dekel14} argue is the result of higher gas
fractions in lower mass galaxies.  Whether galaxies quenches fast through, e.g., violent disk 
instabilities, or slow through halo quenching, more massive galaxies will tend to cross this
quenching threshold earlier in time. 

The predicted redshift evolution of the quenching threshold for central galaxies 
from \citet{Voit15} is shown as a thin black line in Figure~\ref{fig:bestfit_sigma_quenched}.
Their paper presents an argument for self-regulated feedback that links a galaxy’s star-formation 
history directly with the circular velocity of its potential well.  \citet{Voit15} hypothesize 
that feedback leads to quenching when the halo mass reaches a critical value that allows supernovae and/or
AGN to push away the rest of the circumgalactic gas. This critical circular velocity is set to
300 km/s at $z$=2 here, resulting in a quenching threshold that tracks the upper envelope of the 
central density observed for the quiescent population.

In Figure~\ref{fig:bestfit_sigma_quenched}, we show that the
quenching threshold differs from the characteristic turnover value.  This difference may either imply 
that galaxies keep growing their centers after star formation starts actively diminishing. Or, 
that there is significant scatter in this quenching threshold, where galaxies with central densities 
in between the characteristic turnover and the quenching threshold have an intermediate 
probability of being quenched.

Although we observe that a high central density appears to predict quiescence on average, we
caution that this does not necessarily imply causation.  \citet{Lilly16} propose instead that galaxies quench their star 
formation according to empirical probabilistic laws that depend solely on the total mass of the galaxy, 
not on the surface mass density or size. With their simple model they can broadly reproduce all of the 
trends between galaxy structure and sSFR that we observe here, including the evolving quenching threshold. 
\citet{Lilly16} argue that because galaxies form their stars inside-out and passive galaxies will form them at earlier epochs 
(higher redshifts), passive galaxies will always have smaller sizes than their star-forming counterparts of 
the same stellar mass at any given redshift.  Determining the cause of quenching is beyond the scope
of this paper.  But we note that even without causation, we show that the high central density holds 
a unique predictive power in identifying the population of galaxies that, on average, will be quiescent.

\section{Summary}
\label{sec:summary}

The aim of this paper is to connect rest-frame optical measurements of the size-mass and 
density-mass relations with robust measurements of the total specific SFRs from a purely 
empirical standpoint.  We can thereby connect galaxy structure
and star formation to better understand the observed bimodal distribution of galaxies across cosmic time
and the quenching of star formation.
This current study extends the original work by \citet{Franx08} on a smaller field that included 
1155 galaxies at 0.2$<$$z$$<$3.5 to now consider a mass-complete sample of 27,893 galaxies at 0.5$<$$z$$<$2.5.
The sample is selected in five extragalactic fields from the 3D-HST photometric catalogs presented in \citet{Skelton14},
combining high spatial resolution 
\emph{HST} NIR imaging from the CANDELS treasury program with total UV+IR SFRs derived from a 
median stacking analysis of \emph{Spitzer}/MIPS 24$\mu$m imaging.  

The main results presented within this paper are summarized as follows:

\begin{enumerate}

\item We find that 50\% of new stars being formed amongst the overall population occurs in galaxies within $\pm$0.13 dex of the average
size-mass relation.  Extremely compact or extended galaxies do not significantly contribute to the total stellar mass budget.

\item We show a flattening in the size-mass relation of quiescent galaxies at stellar masses
below 10$^{10}$ M$_{\odot}$ at 0.5$<$$z$$<$1.0. These lower mass quiescent galaxies exhibit 
slightly higher sSFRs relative to more massive galaxies at the same epoch, suggesting more recent 
assembly.  However, the 
sSFRs of quiescent galaxies at fixed stellar mass do not show significant variations.

\item After removing the well known correlations between stellar mass and star formation rate 
and galaxy size, we show that star-forming galaxies show a weak dependence of their star formation 
rates on galaxy size.  The residual offset in size for star-forming galaxies in the lowest quartile when
rank ordered by sSFR is 0.27$\pm$0.06 dex smaller than the highest sSFR quartile. Similarly,
when instead rank ordering by the residual size offsets, the smallest galaxies are lower sSFRs by 0.11$\pm$0.02 dex
than that of the largest galaxy quartile. Similar trends are found amongst massive galaxies in 
simulations \citep[e.g.,][]{Furlong15}, though greatly amplified relative to the observations.

\item We find that the independence of star formation rate on galaxy size is not sensitive to 
the timescale on which the star formation rate is probed, with dust-corrected H$\alpha$ sSFRs yielding similar 
trends.

\item We confirm earlier studies \citep[e.g.,][]{Franx08}, showing that the central stellar density is a key parameter connecting 
galaxy morphology and star formation histories: stacks of galaxies with high central densities are red and 
have increasingly lower sSFRs, whereas galaxies stacked with low central densities are blue and 
have a roughly constant (higher) sSFRs at a given redshift interval.  

\item We use a broken power-law to parameterize the correlation between log(sSFR) and central density, log($\rho_{1}$), 
showing remarkably little scatter between the average measurements.  

\item We find strong evolution in the central density threshold for quenching, 
as defined by both a constant and evolving threshold in sSFR, decreasing by $>$0.5 dex from 
$z$$\sim$2 to $z$$\sim$0.7.  Similarly, while the threshold in central circular velocity where most galaxies 
are considered quenched is $>$300 km/s at $z$$\sim$2, this decreases to $\sim$150 km/s by $z$$\sim$0.7.  
\end{enumerate}

We show that neither a high $n$ nor a compact galaxy size
will uniquely predict quiescence, whereas a threshold in central density (or velocity) may be a more robust and unique 
observable signature when considering the overall galaxy population.
However, we emphasize that correlations between structure and star formation do not prove a causal effect. 
For example, it remains to be seen whether small scale structure (at the scale of the stars) or large 
scale parameters (the scale of the dark matter halo) dominate the physical processes which quench galaxies.
While we have presented the average global trends of the sSFR with structural parameters ($r_{e}$, $n$, $\Sigma_{e}$, $\rho_{1}$, 
and $v_{\mathrm{circ},1}$) amongst a mass-complete sample of galaxies using high 
resolution \emph{HST}/WFC3 imaging and deep Spitzer/MIPS 24$\mu$m imaging, future studies with 
the James Webb Space Telescope mid-IR spectroscopic and photometric capabilities will yield
robust measurements of SFR for \emph{individual} galaxies across the star-formation sequence.  
Such studies will allow us to resolve the detailed trends within the star-forming population 
as a function of structure.

\begin{acknowledgements}
We thank the anonymous referee for useful comments and a
careful reading of the paper.
The authors wish to acknowledge Sandy Faber and Mark Voit for helpful discussions.
KEW gratefully acknowledges support by NASA through Hubble Fellowship grant 
\#HST-HF2-51368 awarded by the Space Telescope Science Institute, which is operated 
by the Association of Universities for Research in Astronomy, Inc., for NASA, under 
contract NAS 5-26555. The authors are grateful to the many colleagues who have
provided public data and catalogs in the five deep 3D-HST fields; high redshift galaxy science
has thrived owing to this gracious mindset and the TACs and the Observatory Directors who have encouraged this.
This work is based on observations taken by the 3D-HST Treasury Program (GO 12177 and 12328)
with the NASA/ESA HST, which is operated by the Associations of Universities for Resarch in Astronomy, Inc.,
under NASA contract NAS5-26555.
\end{acknowledgements}

\appendix
\renewcommand{\thefigure}{A\arabic{figure}}
\setcounter{figure}{0}

In order to remove the well known correlations between both star formation rate and galaxy size
with stellar mass and redshift, we fit the best-fit coefficients describing the 
log(SFR)-log(M$_{\star}$) relations in \citet{Whitaker14b} and the log($r_{e}$)-log(M$_{\star}$) 
relation in \citet{vanderWel14}. The methodology
and data itself are identical between the present study and these earlier works.

\begin{figure*}[b]
\leavevmode
\centering
\includegraphics[width=0.85\linewidth]{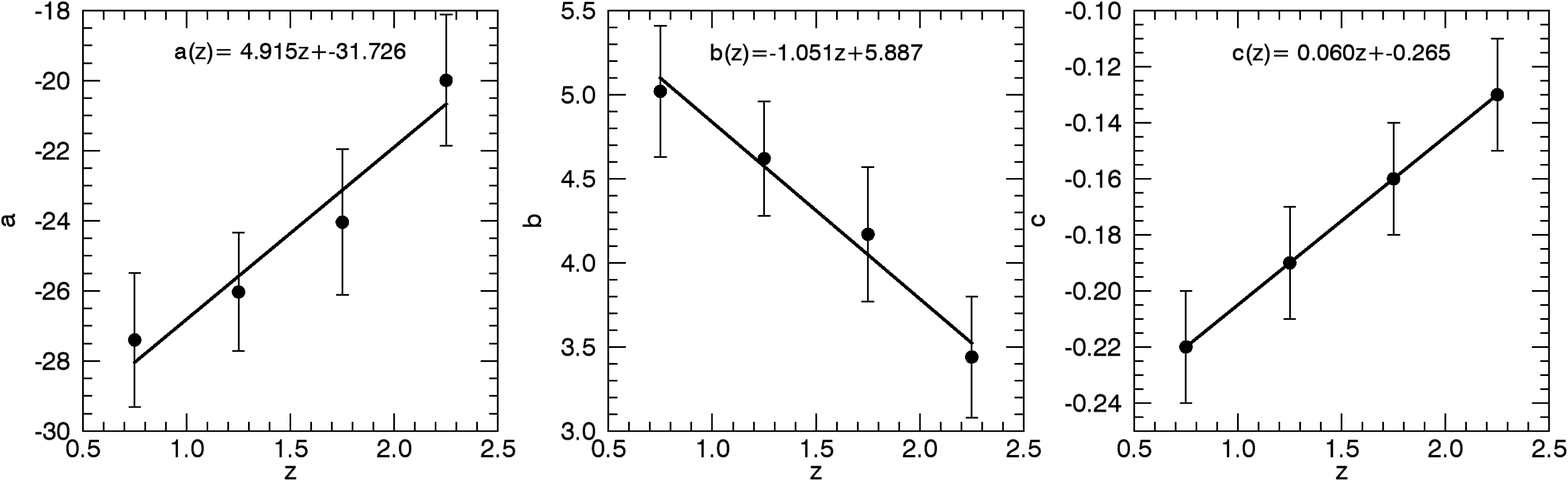}
\caption{Redshift evolution of the best-fit polynomial coefficients to the average observed
logarithmic relation between star formation rate and stellar mass from \citet{Whitaker14b}.
The linear coefficients defining the black solid lines in each panel are listed at the top of each panel.}
\label{fig:sfr_bestfit}
\end{figure*}

\begin{figure*}
\leavevmode
\centering
\includegraphics[width=0.6\linewidth]{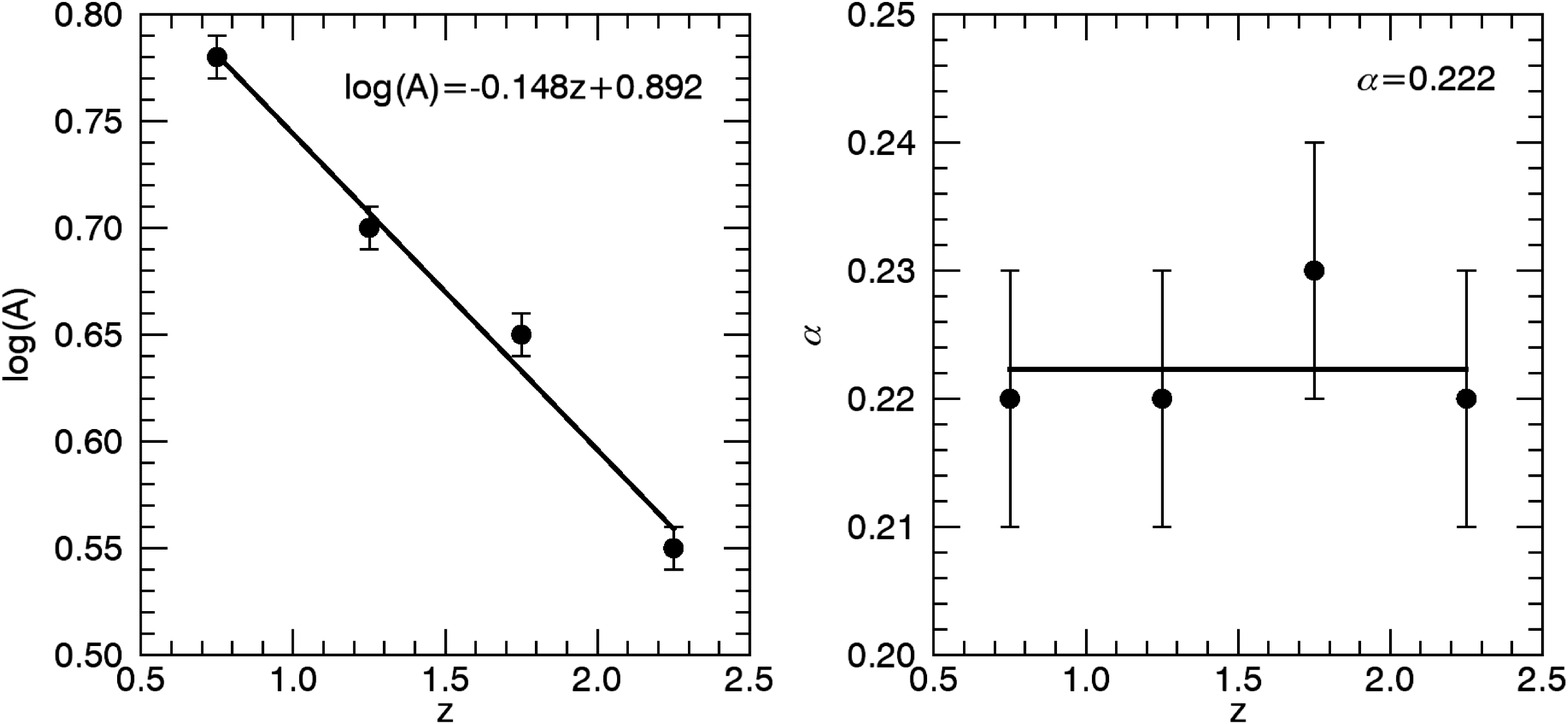}
\caption{Redshift evolution of the best-fit coefficients to the average observed
logarithmic relation between galaxy size and stellar mass from \citet{vanderWel14}.
The linear coefficients defining the black solid lines are listed at the top of each panel.}
\label{fig:size_bestfit}
\end{figure*}

The measured log(SFR)-log(M$_{\star}$) relation is defined in Equation 2 of \citet{Whitaker14b}.
In Figure~\ref{fig:sfr_bestfit}, we fit the redshift evolution of the observed relations
with simple least-square linear fits for the three polynomial coefficients, weighted by their
respective uncertainties.  We therefore parameterize the stellar mass and redshift dependence 
of the star formation sequence as follows,

\begin{equation}
\log \mathrm{SFR}(z,M_{\star}) =  [a_{1}z+a_{2}] + [b_{1}z+b_{2}]\log\left(\frac{M_{\star}}{M_{\odot}}\right) + [c_{1}z+c_{2}]\log\left(\frac{M_{\star}}{M_{\odot}}\right)^{2}
\label{eq:sfrmass}
\end{equation}

\noindent where the best-fit parameters describing the redshift evolution of the polynomial coefficients are
presented at the top of each panel in Figure~\ref{fig:sfr_bestfit}.

In order to remove the average redshift and stellar mass dependence on galaxy size, we also 
subtract the following stellar mass and redshift dependent log($r_{e}$)-log(M$_{\star}$) 
relation from Figure~\ref{fig:size_bestfit}:

\begin{equation}
\log r_{e}(z,M_{\star}) =  [\log(A_{1})z+\log(A_{2})] + [\alpha_{1}z+\alpha_{2}]\times[\log(M_{\star})-\log(5\times10^{10}~M_{\odot})] 
\label{eq:sizemass}
\end{equation}

\noindent with the best-fit redshift evolution of the normalization to the size-mass relation presented
at the top of the left panel in Figure~\ref{fig:size_bestfit}.  The slope of the size-mass relation for star-forming
galaxies only is assumed to be roughly constant, with $\alpha$=0.22. 
Given the average redshift and stellar mass of each bin across
the size-mass plane, we use the two equations above to subtract the average correlations.

\begin{figure*}[t]
\leavevmode
\centering
\includegraphics[width=0.83\linewidth]{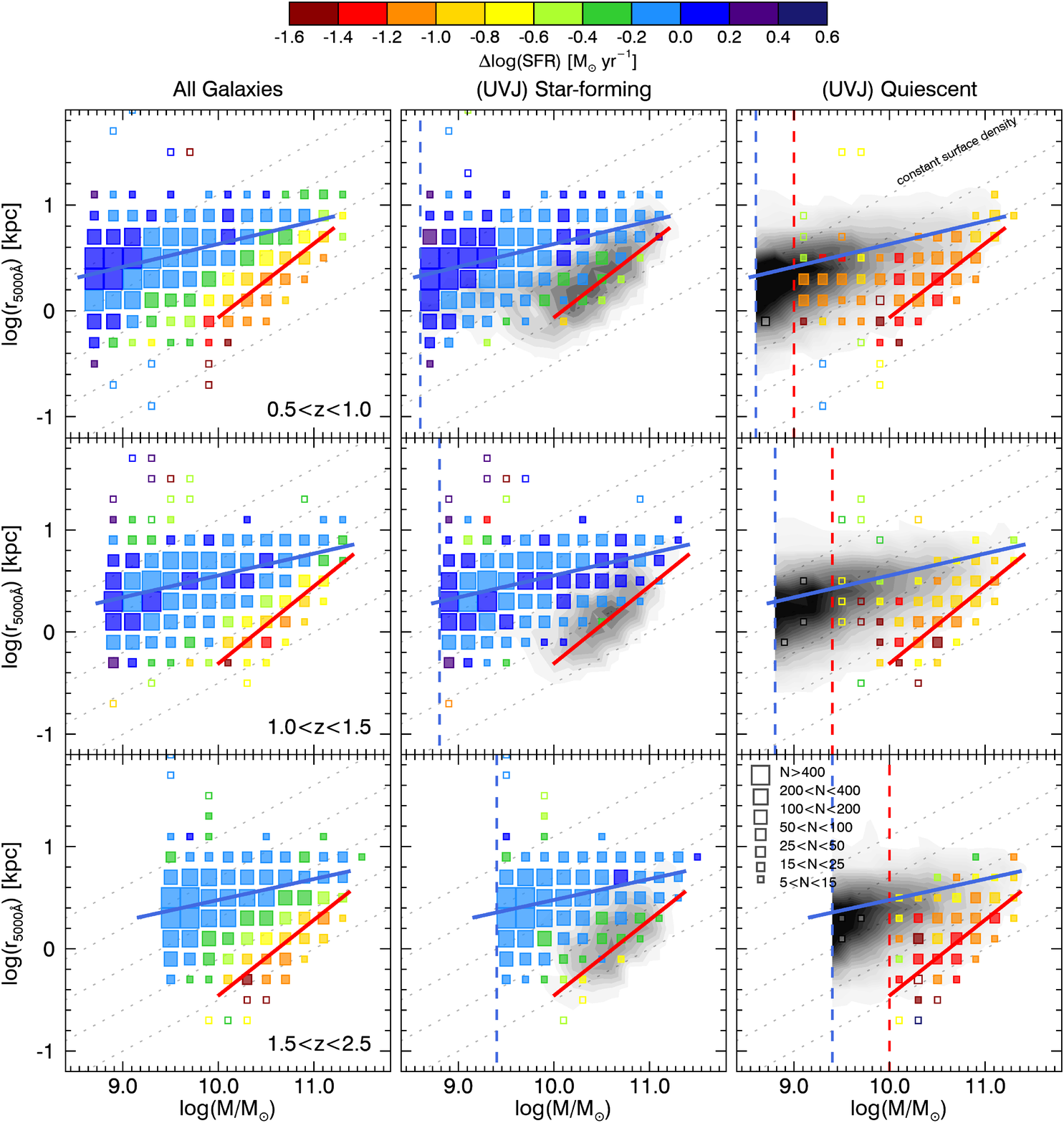}
\caption{Rest-frame 5000\AA\ size of galaxies as a function of stellar mass, color-coded by the logarithmic deviation
from the average log(SFR)-log(M$_{\star}$) relation in 0.2 dex bins.
The size of the symbol depends on the number of galaxies that enter each bin.                         
The vertical dashed lines correspond to the stellar mass limits down to which the structural
parameters can be trusted for star-forming (blue) and
quiescent (red) populations.} 
\label{fig:size2}
\end{figure*}

We present the same data in Figure~\ref{fig:size2} as in Figure~\ref{fig:size}, the rest-frame 
5000\AA\ size of galaxies as a function of their stellar mass in three redshift intervals, but instead
color-code by the logarithmic deviation from the average log(SFR)-log(M$_{\star}$) relation as defined
in Equation~A\ref{eq:sfrmass}.  Whereas we no longer see the turnover towards lower sSFRs in the 
log(SFR)-log(M$_{\star}$) relation at the massive end by definition, we instead highlight this lower envelope of 
compact galaxies with decreased star formation efficiency.  These galaxies are also evident in 
Figure~\ref{fig:size_sfr}, which further takes into account the size-mass relation as defined 
in Equation~A\ref{eq:sizemass}.  From Figure~\ref{fig:size2}, we see that typical larger star-forming galaxies
have SFRs consistent within 1$\sigma$ with the log(SFR)-log(M$_{\star}$) relation at each epoch.  These results
support the main conclusion reached more quantitatively from Figure~\ref{fig:sfuniverse}.

\addcontentsline{toc}{chapter}{\numberline {}{\sc References}}

\end{document}